\documentclass[traditabstract]{aa}
\usepackage{natbib}
\bibpunct{(}{)}{;}{a}{}{,} 
\usepackage{txfonts}
\usepackage{graphicx}
\usepackage{dcolumn}
\newcommand{\kms}{{\,\rm km\,s}^{-1}} 
 
\newcommand{\nodata}{\multicolumn{1}{c}{$\cdots$}}

\def\la{\mathrel{\hbox{\rlap{\hbox{\lower4pt\hbox{$\sim$}}}\hbox{$<$}}}}

\renewcommand{\mag}{\mbox{$\;$mag}}
\def\figsize{0.900} 

\begin{document}
\title{New period-luminosity and period-color\\ relations of classical
       Cepheids}
\subtitle{IV. The low-metallicity galaxies IC\,1613, WLM, Pegasus,
              Sextans A and B, and Leo A in comparison to SMC}

\author{G.~A. Tammann\inst{1} \and B. Reindl\inst{1} \and A. Sandage\inst{2}}

\institute{%
     Department of Physics and Astronomy, Univ. of Basel
     Klingelbergstrasse 82, 4056 Basel, Switzerland \\
     \email{g-a.tammann@unibas.ch}
\and
     The Observatories of the Carnegie Institution of Washington, 
     813 Santa Barbara Street, Pasadena, CA 91101, USA. -- Deceased on
     November 13, 2010
}

\date{Received {21 December 2010} / Accepted {22 April 2011}}

\abstract{%
The metal-poor, fundamental-mode (P0) and first-overtone (P1) Cepheids
in the dwarf galaxies IC\,1613, WLM, Pegasus, Sextans~A, Sextans~B,
and Leo~A are compared with the about equally metal-poor Cepheids of
the Small Magellanic Cloud (SMC). The period-color (P-C) and
period-luminosity (P-L) relations of the seven galaxies are
indistinguishable, but differ distinctly from those in the Large
Magellanic Cloud (LMC) and the solar neighborhood. 
Adopting $(m-M)^{0}_{\rm SMC}=18.93$ from independent evidence, one 
can determine reliable distance moduli for the other dwarf galaxies of
$(m-M)^{0}=24.34\pm0.03$, 
          $24.95\pm0.03$, 
          $24.87\pm0.06$,
          $25.60\pm0.03$ (mean for Sextans~A \&~B), and 
          $24.59\pm0.03$, respectively. 
}

\titlerunning{New P-L and P-C Relations of metal-poor Cepheids}
\maketitle

\section{Introduction}
\label{sec:1}
It has been shown that the character of the period-luminosity (P-L)
relations varies particularly at short wavelengths as a function of the
metallicity (\citealt*{STR:09}, in the following Paper~III).  
This implies the {\em prediction\/} that the very low-metallicity
Cepheids in IC\,1613, WLM, and the Pegasus dwarf system should follow
P-L relations that are more similar to those of the SMC than those
defined by the more metal-rich Cepheids in the LMC and the Galaxy. The
same prediction holds for the period-color (P-C) relations. The
purpose of this paper is therefore to compare the P-L and P-C
relations of the three above-mentioned galaxies with the
corresponding, well-defined relations of the SMC (\citeauthor{STR:09},
but revised here in Sect.~\ref{sec:2}). Fundamental-mode (P0) as
well as first-overtone (P1) Cepheids are considered. 

     In addition, we consider the Cepheids in Sextans~A and Sextans~B
(joined here into one set) and in Leo~A. The metallicity of the young
population in these galaxies is still lower by a factor of three to four
than in SMC. The question is whether this additional underabundance
has a noticeable effect on the P-L and P-C relations.  

     The most metal-poor galaxy known, i.e.\ I\,ZW\,18 with
[O/H]$_{T_{\rm e}}=7.2$ \citep{Skillman:Kennicutt:93}, is not
considered here because so far only a single Cepheid is known in the
useful period range \citep{Fiorentino:etal:10}.

     The metallicities in the $T_{e}$-based system of
\citet{Zaritsky:etal:94} of the galaxies in the present sample and
their Galactic foreground reddenings \citep[from][]{Schlegel:etal:98}
are given in Table~\ref{tab:01}. All data are corrected in the
following for foreground reddening and absorption.  

\begin{table*}
\centering
\caption{Metallicities and foreground reddening of the sample galaxies.}
\label{tab:01}
\small
\begin{tabular}{llll}
\hline
\hline
\noalign{\smallskip}
   &
   \multicolumn{1}{c}{[O/H]$_{T_{\rm e}}$}    &
   \multicolumn{1}{c}{Source}                 &
   \multicolumn{1}{c}{$E(B\!-\!V)_{\rm Gal}$} \\
\noalign{\smallskip}
\hline
\noalign{\smallskip}
SMC            & 7.98 & \citealt{Sakai:etal:04}                   & variable     \\ 
IC\,1613       & 7.86 & \citealt{Sakai:etal:04}                   & 0.025        \\ 
WLM            & 7.74 & \citealt{Sakai:etal:04}                   & 0.037        \\ 
Pegasus        & 7.92 & \citealt{Skillman:etal:97}                & 0.066        \\
Sex A \& B     & 7.52 & \citealt{Skillman:etal:89}                & 0.044, 0.032 \\
Leo A          & 7.38 & \citealt{Skillman:etal:89,vanZee:etal:06} & 0.021        \\
\noalign{\smallskip}
\hline
\end{tabular}
\end{table*}

     The P-C and P-L relations of the P0 and P1 Cepheids in the five
sample galaxies and their distances are discussed in
Sects.~\ref{sec:3}-\ref{sec:7}. The mean P0 and P1 distances are
discussed in the light of independent distance determinations in
Sect.~\ref{sec:8}. In Sect.~\ref{sec:9}, we compare the P-C and P-L
relations of the metal-poor sample galaxies with the corresponding
relations for more metal-rich Cepheids.

\section{The P-C and P-L relations of SMC as templets for
  fundamental-mode (P0) and first-overtone (P1) pulsators} 
\label{sec:2}
The P-C and P-L relations of the P0 Cepheids in SMC were derived in
\citeauthor{STR:09} using the $B$, $V$, $I$ photometry of the OGLE
program \citep{Udalski:etal:99b}. Following common practice, the
Cepheids with $\log P<0.4$ were excluded. The remaining about 450
Cepheids are individually corrected for internal absorption by
\citeauthor{Udalski:etal:99b}. By performing fits to the data, we were
unable to unambiguously decide whether the relations have a break at
$\log P=1.0$ as in LMC or not. The (P-C) relation in $(B\!-\!V)$
showed the break clearly, whereas in $(V\!-\!I)$ and the P-L relations
it remained insignificant. 
        
     The analysis is repeated here by fitting two linear regressions
to the total of about 1100 P0 Cepheids (including $\log P<0.4$) and
treating the position of the break as a free parameter. The additional
requirement is a minimum discontinuity of the P-C and P-L relations at
the break period. Higher-order regressions are, of course, possible,
but the linear fits are adequate for all practical purposes and
facilitate the comparison with other galaxies.

     In addition, the corresponding relations in $(B\!-\!V)$,
$(V\!-\!I)$, and $B$, $V$, and $I$ are derived in Sect.~\ref{sec:2:2}
also for the P1 Cepheids of SMC as identified by
\citet{Udalski:etal:99b}. 

     The very large set of SMC Cepheids with $V$ and $I$ photometry by
\citet{Soszynski:etal:10} defines P-C and P-L relations whose slopes
agree with the ones derived here, but the magnitudes are not corrected
for variable internal absorption. Therefore, the data were not used
here.

\subsection{Fundamental-mode (P0) P-C and P-L relations of SMC}
\label{sec:2:1}
\begin{figure*}
   \centering
   \resizebox{\figsize\hsize}{!}{\includegraphics{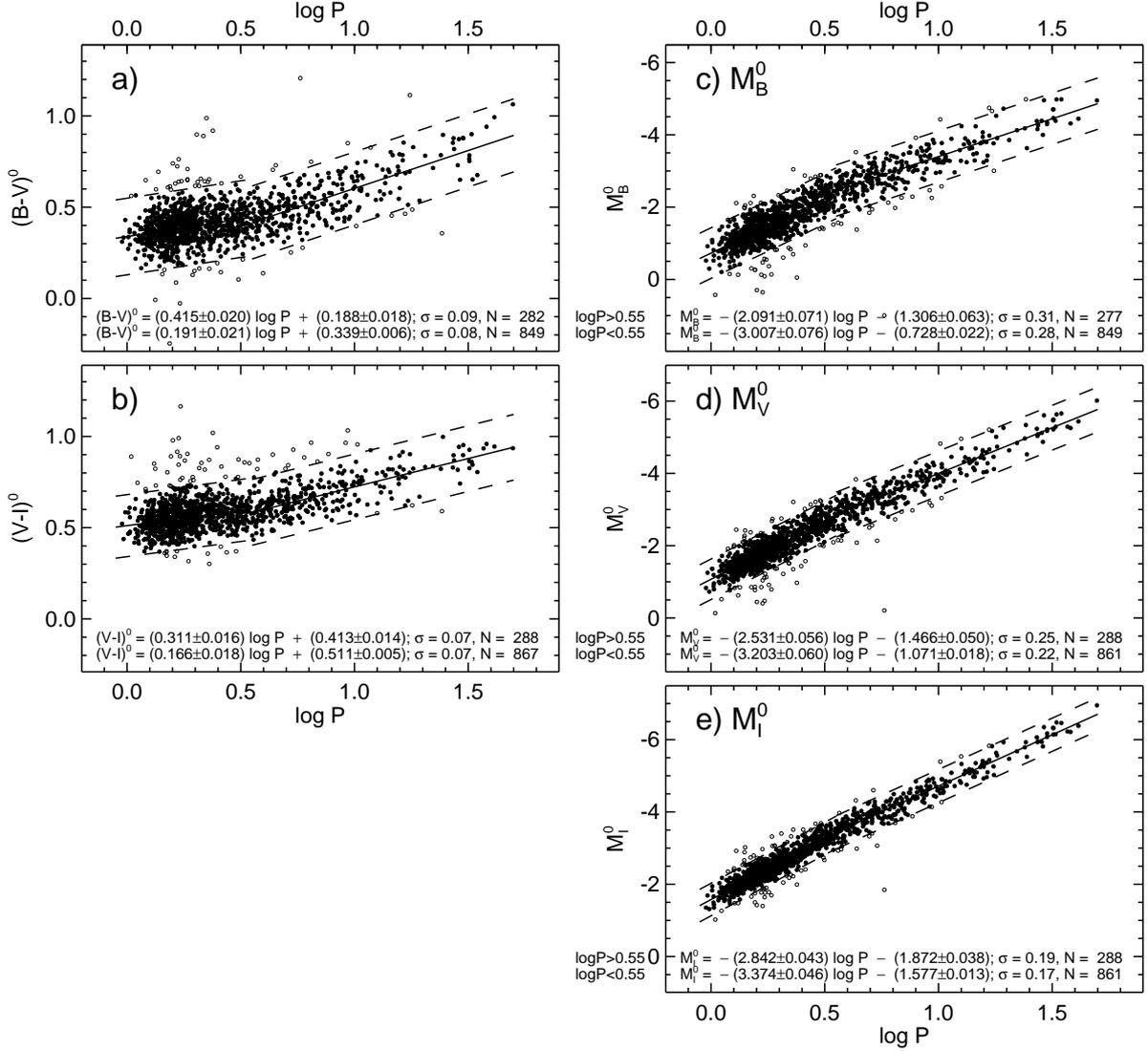}}
   \caption{The revised P-C and P-L relations of the P0 Cepheids of
     SMC including the shortest-periods and with the adopted break at
     $log P=0.55$. The dashed lines are the  $2\sigma$ boundaries;
     objects outside are excluded. a) \& b) The P-C relations in
     $(B\!-\!V)$ and $(V\!-\!I)$, respectively. c) - e) The P-L
     relations in $B$, $V$, and $I$, respectively. 
}
    \label{fig:01}
\end{figure*}

The P-C relations in $(B\!-\!V)$ and $(V\!-\!I)$ of the P0 Cepheids of
SMC are shown in Fig. \ref{fig:01}a \& b. They are well-defined for
Cepheids with about $0.0 < \log P < 1.5$ and show a striking break
near $\log P=0.55$. At this point, the two regressions merge into each
other with little discontinuity. The remaining mismatch is smaller
than the statistical errors of the two segments.

     The P-L relations in $B$, $V$, and $I$ are shown in
Fig.~\ref{fig:01}c to e. Their highly significant break is found near
$\log P=0.55$ in agreement with the P-C relations. For the calibration
of the P-L relations, the distance of SMC of $(m-M)_{\rm SMC}^{0}=18.93$
is adopted from \citet[][hereafter TSR\,08a, Table~7]{TSR:08a}.
This value with an estimated uncertainty of $<\!0.1\mag$ is the mean
of different distance determinations, but is {\em independent\/} of
the P-L relation of Cepheids. All distances in this paper are based on
this zero-point. 

     The equations of the P-C and P-L relations, after $2\sigma$
clipping, are shown at the bottom of the respective panels in
Fig.~\ref{fig:01} for the Cepheids below and above the break point;
they are repeated in Table~\ref{tab:05} below. 

\subsection{First-overtone (P1) P-C and P-L relations of SMC}
\label{sec:2:2}
\begin{figure*}
   \centering
   \resizebox{\figsize\hsize}{!}{\includegraphics{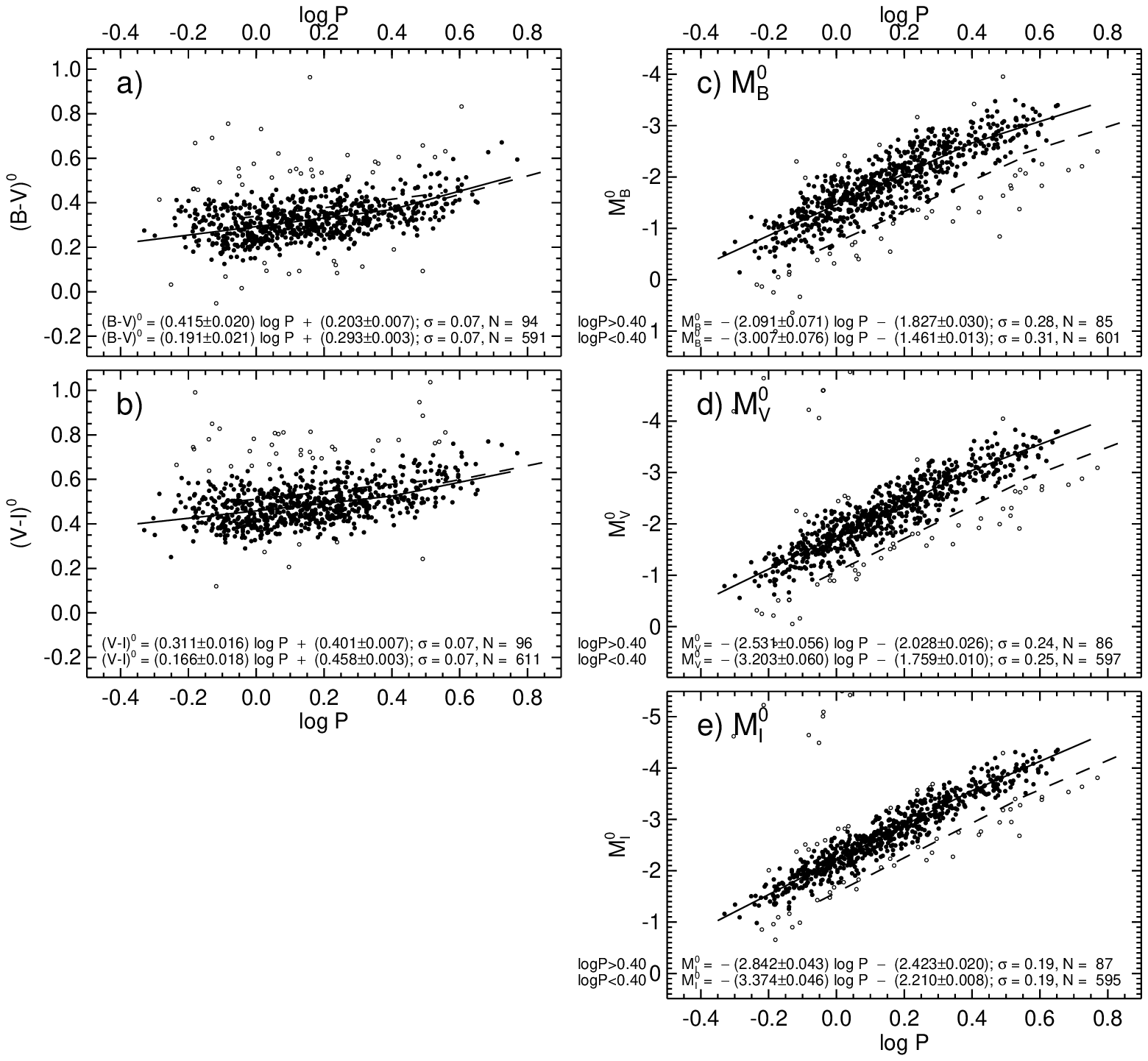}}
   \caption{The P-C  and P-L relations of the P1 Cepheids of SMC with
     the adopted break at $\log P=0.4$. The dashed lines show the
     locus of the P0 Cepheids. 
}
    \label{fig:02}
\end{figure*}

The P1 Cepheids of SMC span an interval of $-0.3 < \log P < 0.6$. They
were identified as P1 pulsators by \citet{Udalski:etal:99b}, but the
sample still contains a few rather faint variables that are probably
P0 Cepheids. The clearly broken P-C and P-L relations, corrected as
before for variable internal absorption, are shown in
Fig.~\ref{fig:02}. They are adequately fit by two linear regressions 
with different slopes. The slopes, even for different positions of the
break point, are so similar to the slopes of the two segments of the
P0 Cepheids that it is assumed that corresponding segments are
parallel. In that case, the P-C and P-L relations yield a good
compromise break at $\log P=0.4$. 
The equations of the different relations are given in Table~\ref{tab:02}
following the scheme $x = a\log P + b$. 

\begin{table*}
\centering
\caption{P-C and P-L relations of first-overtone (P1) Cepheids in SMC.}
\label{tab:02}
\small
\begin{tabular}{lrrcrr}
\hline
\hline
\noalign{\smallskip}
   &
   \multicolumn{2}{c}{$\log P<0.4$} & &
   \multicolumn{2}{c}{$\log P>0.4$} \\
\noalign{\smallskip}
\hline
\noalign{\smallskip}
$(B\!-\!V)^{0}$ & $ 0.191\pm0.021$ & $ 0.293\pm0.003$ & & $ 0.415\pm0.020$ & $ 0.203\pm0.007$ \\ 
$(V\!-\!I)^{0}$ & $ 0.166\pm0.018$ & $ 0.458\pm0.003$ & & $ 0.311\pm0.016$ & $ 0.401\pm0.007$ \\[2pt]
$M^{0}_{B}$     & $-3.007\pm0.076$ & $-1.461\pm0.013$ & & $-2.091\pm0.071$ & $-1.827\pm0.030$ \\  
$M^{0}_{V}$     & $-3.203\pm0.060$ & $-1.759\pm0.010$ & & $-2.531\pm0.056$ & $-2.028\pm0.026$ \\
$M^{0}_{I}$     & $-3.374\pm0.046$ & $-2.210\pm0.008$ & & $-2.842\pm0.043$ & $-2.423\pm0.020$ \\
\noalign{\smallskip}
\hline
\end{tabular}
\end{table*}

     The relative position of the P-L relation of fundamental-mode and
first-overtone Cepheids suggests a period ratio at constant luminosity
of metal-poor Cepheids of P0/P1$=1.4$. 

     We use the occasion to also revise here the P-C and P-L relations
of LMC. They were derived in \citeauthor{STR:04} (\citealt*{STR:04})
from a sample of 634 Cepheids using $B$, $V$, and $I$ photometry by
\citet{Udalski:etal:99a} and other sources. The relations showed a
highly significant break at $\log P=1.0$, but about 100 Cepheids with
$\log P < 0.4$ were excluded. With the inclusion of these objects, the
best-fit break point is shifted to $\log P=0.9$. The corresponding
P-L relations are given in Table~\ref{tab:05} below.  

\section{IC\,1613}
\label{sec:3}
The first 27 Cepheids in IC\,1613 were found by W.~Baade. He did not
publish them because they defined a P-L relation much flatter than in
LMC, which he suspected could be caused by a scale error in his
photographic mpg magnitudes. \citet{Sandage:71}, after fitting the
magnitudes into a photoelectric $B$ scale, confirmed the flat slope,
but showed that the deviations from the LMC slope could be explained
by the intrinsic width of the instability strip and a statistical
fluke. \citet{Freedman:88} determined CCD magnitudes in the UBV system
for nine of Baade's Cepheids, but their number is too small to provide
a reliable slope; she fitted them to the LMC P-L relations, but a
flatter slope fits the data at least as well.  
\citet{Udalski:etal:01} provided mean $VI$ magnitudes for many
Cepheids in IC\,1613 and found no significant deviations from the LMC
slope known at that time. Yet additional $BVI$ photometry of the
Cepheids in IC\,1613 by \citet{Antonello:etal:06} reopened the question 
of the agreement between IC 1613 and LMC, a question that gained
new weight after the P-L and P-C relations of LMC were shown to
display a pronounced break at $\log P=1.0$ 
(\citealt{Tammann:Reindl:02}; \citealt{Tammann:etal:02};
\citealt*{STR:04}, \citeauthor{STR:04};
see also
\citealt{Kanbur:Ngeow:04}; \citealt{Ngeow:etal:05};
\citealt{Koen:Siluyele:07}; \citealt{Kanbur:etal:07}). 
Additional Cepheids in IC\,1613 were found by \citet{Bernard:etal:10},
which have short or very short periods and are useful for the
definition of the tails of the P-L relations.  

     Several authors have obtained photometry of IC\,1613 Cepheids in
the near- or mid-infrared. These cannot, however, be compared
with LMC or SMC, either because the number of Cepheids is too small or
the corresponding data are missing in the Clouds.

     The Cepheids of IC\,1613 are here compared with those of
{\em SMC}. This is because the two galaxies have very similar
metallicities and as a consequence of this a significant population of
very short-period Cepheids, in contrast to LMC.

\subsection{The data}
\label{sec:3:1}
The following Cepheid data were used to define the P-L and P-C
relations of IC\,1613:
\begin{enumerate}
\item
   \citet{Udalski:etal:01} obtained mean $V$ and $I$ photometry
   for 138 Cepheids in the framework of the extensive OGLE II project
   (78 with $\log P> 0.4$).
   They excluded the overtone pulsators, two type II Cepheids,
   two blends, and the outlyer 13682 (= V39 from \citealt{Sandage:71}).
   We exclude in addition the Cepheids with $\log P<0.4$ because their
   separation into fundamental and overtone pulsators is ambiguous.
   This leaves 60 fundamental pulsators in the sample. 
\item
   \citet{Antonello:etal:06} published mean $BVRI$ magnitudes of
   49 P0 Cepheids from the sample of \citet{Udalski:etal:01} and 3
   additional ones. They observed them at only a few epochs, but
   used the known light curves in $V$ and $I$ to also construct
   the light curves in $B$ and $R$ following the method of
   \citet{Freedman:88}. They provided mean $B$ magnitudes for 49, $V$
   magnitudes for 52, and $I$ magnitudes for 51 Cepheids. (The $R$
   magnitudes are not considered here).
   Six of the 48 Cepheids with three-color photometry fall outside the
   boundary defined by SMC in the $(B\!-\!V)$ versus $(V\!-\!I)$
   diagram (see Fig.~\ref{fig:02}a below); these objects are excluded. 
\item
   \citet{Dolphin:etal:01} added five P0 Cepheids with
   $V$ and $I$ photometry.
\item
   \citet{Bernard:etal:10} presented ACS HST photometry of many
   faint variables in IC\,1613 including 49 Cepheids.
   Twenty-six Cepheids, all with $VI$ photometry, are identified by the
   authors as P0 pulsators, 12 of which are also in the sample of
   \citet{Udalski:etal:01}.
   Of the remaining 23 Cepheids, 16 are classified as first-overtone
   (P1) pulsators.\\ 
   We tentatively added the 147-day Cepheid V22 with $BVI$
   photometry by \citet{Freedman:88}; the photometry should be
   confirmed because other Cepheids from the same source agree poorly
   with \citet{Antonello:etal:06}. 
\end{enumerate}

\begin{figure*}
   \centering
   \resizebox{\figsize\hsize}{!}{\includegraphics{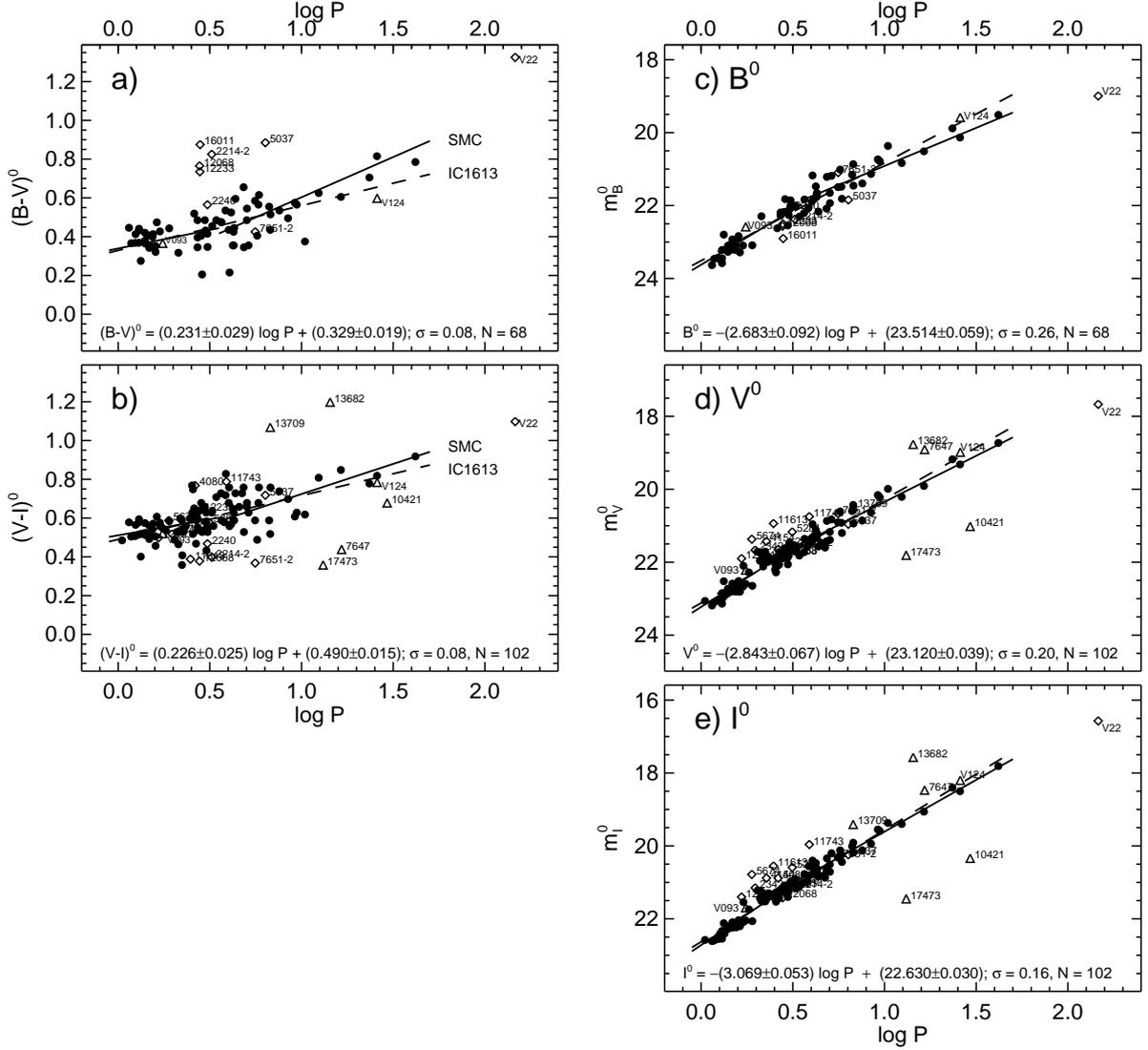}}
   \caption{The P-C and P-L relations of the fundamental-mode (P0)
   Cepheids of IC\,1613. The full lines are the corresponding
   relations of SMC. Forced linear fits over the whole period interval
   are shown as dashed lines (their equations are given at the bottom
   of each panel). The agreement between IC 1613 and SMC is striking.
   Cepheids excluded by the original authors (triangles) or by ourselves
   (diamonds) are shown as open symbols.}
   \label{fig:03}
\end{figure*}

The four samples were merged resulting in a sample of 124 P0 Cepheids,
of which 22 were excluded as outliers (marked in Fig.~\ref{fig:03}) in
addition to those already discussed. Mean magnitudes from more than
one source were averaged, as also done for the following
galaxies. The final sample consists of 102 P0 Cepheids with $V$
and $I$ photometry, of which 68 have also $B$ magnitudes. The 16 P1
Cepheids have $B$, $V$, and $I$ magnitudes.

\subsection{The P-C and P-L relations of IC\,1613}
\label{sec:3:2}
The P-C relations in $(B\!-\!V)^{0}$ and $(V\!-\!I)^{0}$ and the P-L
relations in $B$, $V$, and $I$ of the P0 Cepheids and the excluded
variables are shown in the five panels of Fig.~\ref{fig:01}.

     The P-C and P-L relations in panels a--e are described well by the 
broken P-C and P-L relations of SMC adopted in Sect.~\ref{sec:2}
(full lines). In the case of the P-L relations, SMC was shifted in
apparent magnitude to achieve the best fit.
The break at $\log P=0.55$ is clearly visible in all five panels.
The scatter in the data for the IC\,1613 Cepheids about the SMC
relations in the five panels is the same as of the SMC Cepheids
themselves. Forced linear fits over the whole period range for the
IC\,1613 and SMC Cepheids agree to within $1\sigma$.
The relations of the two galaxies are indistinguishable.

\subsection{Derived parameters of IC\,1613: reddening and distance}
\label{sec:3:3}
\begin{figure*}
   \centering
   \resizebox{\figsize\hsize}{!}{\includegraphics{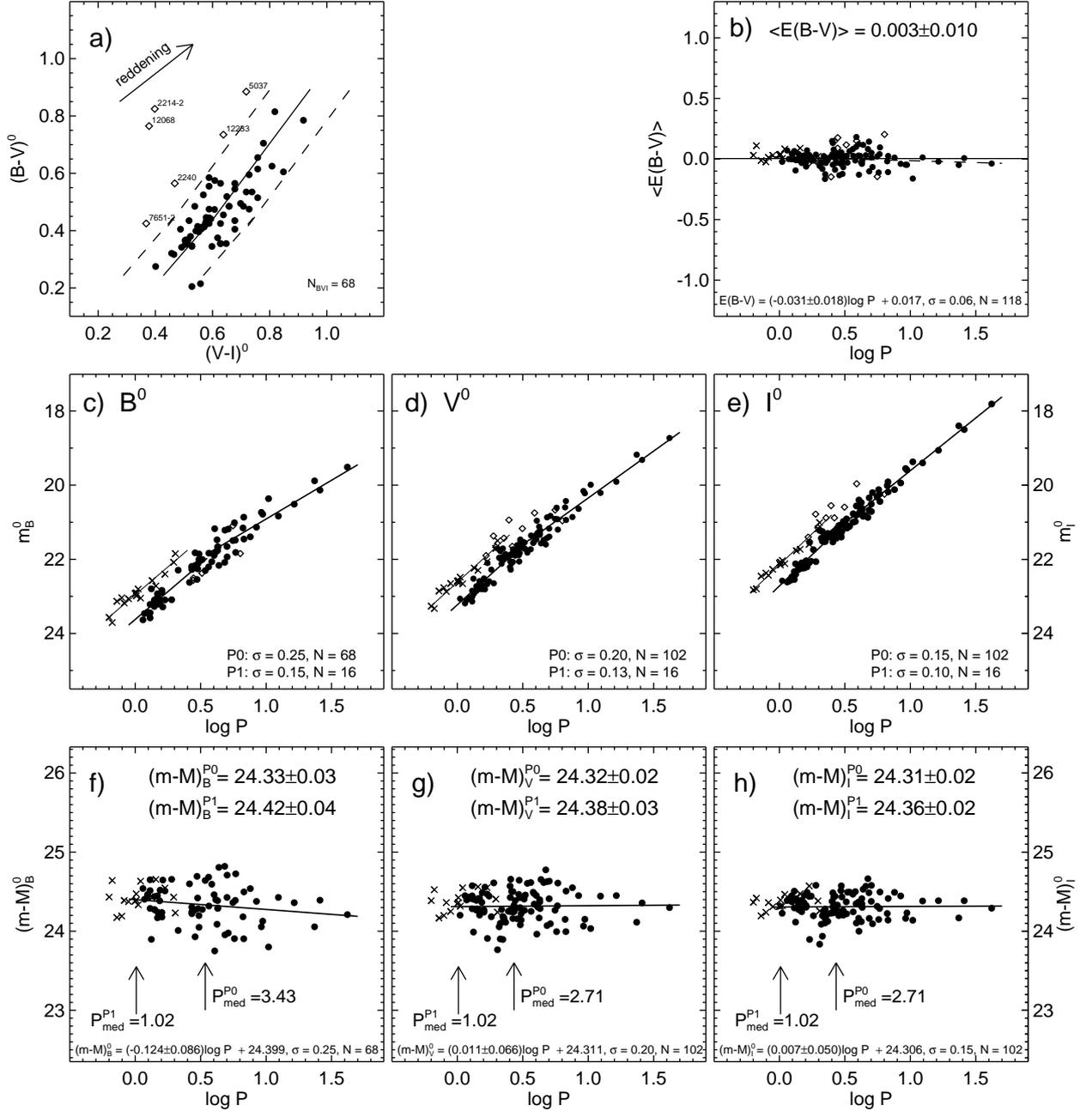}}
   \caption{a) The two-color index diagram $(B\!-\!V)^{0}$ versus
     $(V\!-\!I)^{0}$ of the P0 Cepheids of IC\,1613.
     The region of SMC Cepheids is outlined. Cepheids outside this
     region are identified and shown as open diamonds. They are
     repeated in the following panels, but not used for the fits.
     b) The individual internal color excesses $E(B\!-\!V)$ and
     $E(V\!-\!I)$, combined here into mean values of
     $\langle E(B\!-\!V)\rangle$, as a function of $\log P$. 
     c--e) The cleaned P-L relations in $B$, $V$, and $I$
     of the P0 Cepheids (dots) and P1 (crosses) Cepheids.
     The full and thin lines are the SMC P-L relations for P0 and P1
     pulsators, respectively, fitted to the data. 
     f--h) The individual true distance moduli of the Cepheids of
     IC\,1613 as a function of $\log P$. The full lines are fits to
     only the P0 Cepheids; their equations are indicated at the bottom
     of the panels. The mean distances, read at the median period 
     $P_{\rm med}$, are shown for P0 and P1 Cepheids for each color.}
   \label{fig:04}
\end{figure*}

The Cepheids with three-color photometry are plotted in a two-color
diagram $(B\!-\!V)^{0}$ versus $(V\!-\!I)^{0}$ in Fig.~\ref{fig:04}a,
where the region defined by SMC Cepheids is also shown.
The majority of the IC\,1613 Cepheids lie within the SMC boundaries
except six, which are identified in the figure. These stars are
possible P1 pulsators (cf. their position in Fig.~\ref{fig:04}e), but
are excluded in the following discussion. 

     A comparison of the colors $(B\!-\!V)^{0}$ and $(V\!-\!I)^{0}$ of
the P0 and P1 Cepheids of IC\,1613 with the P-C relation of SMC leads
to their individual color excesses $E(B\!-\!V)$ and $E(V\!-\!I)$.
The latter are transformed into $E(B\!-\!V)$ 
(we note that $E(V\!-\!I)=1.28E(B\!-\!V)$).
The mean values of $E(B\!-\!V)$s are adopted and plotted against
period in Fig.~\ref{fig:04}b. The slight period dependence is
neglected. The scatter in the individual excesses of
$\sigma E(B\!-\!V)=0.06$, which is smaller than in SMC (0.09),
is due to the intrinsic color width of the instability strip.
The overall mean internal excess of $E(B\!-\!V)=0.003\pm0.010$ is
negligible. The tacit assumption is that the intrinsic color of
Cepheids in IC\,1613 is the same as in SMC in agreement with their
similar metallicities. If IC\,1613 Cepheids were intrinsically bluer,
their blueness would have to be closely compensated for by the
corresponding amount of internal absorption, which seems far-fetched.

     The P-L relations in ${B}^{0}$, ${V}^{0}$, and ${I}^{0}$
of the P0 and P1 Cepheids are shown in Fig.~\ref{fig:04}c to e. 
The IC\,1613 data fit the SMC relations very well, their scatter being
the same as for the SMC Cepheids proper.
On the assumption that P0 and P1 Cepheids in IC\,1613 and SMC have the
same luminosity and that $(m-M)_{\rm SMC}^{0}=18.93$, the individual
distances of IC\,1613 Cepheids are derived by comparing them with the
corresponding P-L relations of SMC from Sect.~\ref{sec:2}.
The resulting distances are plotted against period in
Fig.~\ref{fig:04}f to h.  
A certain problem arises from the $B$ data yielding distances that
depend (mildly) on period. This general problem is addressed in
Sect.~\ref{sec:8}. The mean distances, derived from each color, read
at the median period, are shown in the corresponding panels of
Fig.~\ref{fig:04}. If averaged over all three colors, the P0 and P1
Cepheids are found to correspond to  
$(m-M)^{0}=24.32\pm0.02$ and
          $24.38\pm0.03$, respectively. We adopt a number-weighted
overall mean of $24.34\pm0.03$.

     Cepheid distances of IC\,1613 have been determined by different
authors. A selection is compiled in Table~\ref{tab:03} in
chronological order. Distances that are based on assuming the LMC as
their zero-point are normalized here to $(m-M)_{\rm LMC}^{0}=18.52$
(\citeauthor*{TSR:08a}, Table~6), those based on SMC are normalized to
$(m-M)_{\rm SMC}^{0}=18.93$. At first sight, the distances, even if
based on the LMC, agree to within $\pm0.1\mag$. 
However, the agreement of the $E(B\!-\!V)$ values in column~2 is poor
in some cases. Total excesses of  $E(B\!-\!V)>0.06$ (i.e.\ $0.035\mag$
in excess of the Galactic contribution) imply that the IC\,1613 Cepheids
are significantly bluer in $(B\!-\!V)$ than even those in SMC, which
invalidates the proposed use of the relatively red LMC Cepheids as a
template.

\begin{table}[t]
\begin{center}
\caption{Cepheid distances of IC\,1613.}
\label{tab:03}
\tiny
\begin{tabular}{cD..{3}lcl}
\hline
\hline
\noalign{\smallskip}
   \multicolumn{1}{c}{$(m-M)^{0}$}  &
   \multicolumn{1}{c}{$E(B\!-\!V)$} &
   \multicolumn{1}{c}{Cal.}         &
   \multicolumn{1}{c}{passband}     &
   \multicolumn{1}{c}{Source}       \\
   \multicolumn{1}{c}{(1)}          &
   \multicolumn{1}{c}{(2)}          &
   \multicolumn{1}{c}{(3)}          &
   \multicolumn{1}{c}{(4)}          &
   \multicolumn{1}{c}{(5)}          \\
\noalign{\smallskip}
\hline
\noalign{\smallskip}
$24.55$        & 0.03                           & Gal. & $m_{\rm pg}$     & \citealt{Sandage:71}          \\
$24.11\pm0.25$ & 0.05                           & LMC  & $B$              & \citealt{deVaucouleurs:78}    \\
$24.31\pm0.11$ & 0.03                           & Gal. & $H$              & \citealt{McAlary:etal:84}     \\
$24.29\pm0.11$ & \multicolumn{1}{c}{---}        & LMC  & $BVRIH$          & \citealt{Freedman:88b}        \\
$24.41\pm0.14$ & \multicolumn{1}{c}{$\sim$0.04} & LMC  & $BV(W)$          & "                             \\
$24.44\pm0.13$ & 0.03                           & LMC  & $BVRI$           & \citealt{Madore:Freedman:91}  \\
$24.52\pm0.10$ & (0.07)                         & LMC  & $VI(W)$          & \citealt{Macri:etal:01}       \\
$24.45\pm0.07$ & 0.025                          & LMC  & $VI(W)$          & \citealt{Udalski:etal:01}     \\
$24.44\pm0.13$ & \multicolumn{1}{c}{---}        & SMC  & $VI(W)$          & \citealt{Dolphin:etal:03}     \\
$24.31\pm0.04$ & 0.09                           & LMC  & $JK$             & \citealt{Pietrzynski:etal:06} \\
$24.25\pm0.20$ & 0.07:                          & LMC  & $BVRI$           & \citealt{Antonello:etal:06}   \\
$24.48\pm0.12$ & 0.024                          & LMC  & $BVRI$           & "                             \\
$24.32\pm0.02$ & 0.025                          & SMC  & $BVI$            & \citeauthor*{TSR:08a}         \\
$24.29\pm0.12$ & 0.08                           & LMC  & 3.6; 4.5\,$\mu$m & \citealt{Freedman:etal:09}    \\
$24.30\pm0.07$ & \multicolumn{1}{c}{---}        & LMC  & 3.6; 4.5\,$\mu$m & \citealt{Ngeow:etal:09}       \\ 
$24.34\pm0.03$ & 0.025                          & SMC  & $BVI$            & present paper                 \\
\noalign{\smallskip}
\hline
\end{tabular}
\end{center}
\end{table}

     Four entries in Table~\ref{tab:03}, designated with a (W) in
column~4, have above average distances. They were determined by
means of so-called Wesenheit magnitudes $m_{W}$, which are defined as 
$m_{W}(V)= m^{\rm obs}_{V}-R_{V}(B\!-\!V)^{\rm obs}$ or 
$m_{W}(I)= m^{\rm obs}_{I}-R_{I}(V\!-\!I)^{\rm obs}$, where $R$ is the
ratio of total to selective absorption. These pseudomagnitudes were
originally introduced by \citet{vandenBergh:68} and have
been widely used since to account for absorption in an approximate way.
However, the method is only applicable to Cepheids with identical P-C
relations, i.e.\ of the same metallicity. In the case of different 
{\em intrinsic\/} colors, not only the reddening but also the intrinsic
color difference is multiplied with the value $R$, which leads to
systematic distance errors.

\section{WLM}
\label{sec:4}
\citet{SC:85a} found the first 15 Cepheids in WLM, all of which have
periods shorter than $10^{\rm d}$, and provided their light curves in
$B$. For five of them, \citet{Valcheva:etal:07} determined $J$
magnitudes. \citet{Pietrzynski:etal:07} published the data of 59 Cepheids
with $V$ and $I$ magnitudes, which include data for 14 of the Cepheids by
\citeauthor{SC:85a}. 
\citet{Gieren:etal:08} added $J$ and $K$ magnitudes for 31 of
\citeauthor*{Pietrzynski:etal:07} Cepheids.
No P1 Cepheids had previously been identified in WLM.

\begin{figure*}
   \centering
   \resizebox{\figsize\hsize}{!}{\includegraphics{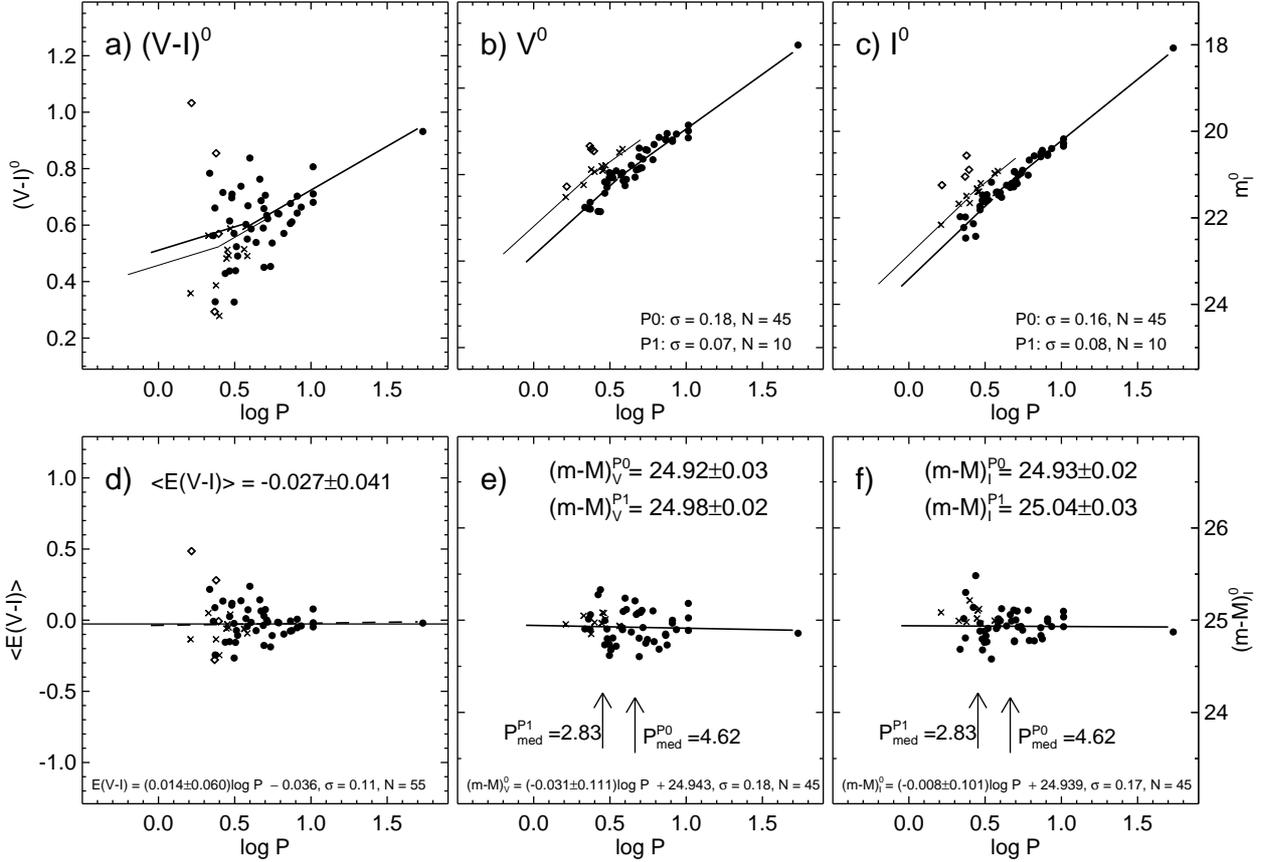}}
   \caption{
   a) The P-C relation in $(V\!-\!I)$ of WLM. Full dots are P0 Cepheids
   throughout, crosses P1 Cepheids, and open diamonds excluded
   variables. The thick line is the P-C relation of SMC for P0, the
   thin line for P1 Cepheids. --
   b) \& c) The P-L relations in $V$ and $I$ of the same objects as in
   a). The full lines are the P-L relations of the P0 (thick line) and
   P1 (thin line) Cepheids in SMC, shifted in magnitude to fit the
   WLM data. The similarity of the WLM and SMC Cepheids is apparent. --
   d) The internal color excesses $E(V\!-\!I)$ of the P0 and P1
   Cepheids in WLM as a function of $\log P$ inferred from a
   comparison with the adopted SMC templets. --
   e) \& f) The individual true distance moduli of the P0 and P1
   Cepheids as a function of $\log P$ inferred from a comparison of
   the true apparent $V$ and $I$ magnitudes with the corresponding SMC
   templets. The full line is a fit to only the P0 Cepheids; their
   equations are indicated at the bottom of the panels. The mean
   distances are read at the indicated median period $P_{\rm med}$.
}
    \label{fig:05}
\end{figure*}

     The P-C and P-L relations of the Cepheids by
\citeauthor{Pietrzynski:etal:07}, corrected for foreground reddening
of $E(V\!-\!I)=0.047$ \citep{Schlegel:etal:98}, are exhibited in
Fig.~\ref{fig:05}a to c where the corresponding relations of SMC are
drawn as full lines. Three objects are excluded by the authors as
being blends or overtone pulsators; one additional outlier (cep55) was
excluded by us. Of the remaining 55 Cepheids, 45 are accepted as P0
Cepheids, the remainder very likely being P1 pulsators. With this
interpretation, the match of WLM with SMC becomes impressively good.
The scatter of the points about the SMC templets is almost the same
as that of the SMC Cepheids.

     The comparison of the $(V\!-\!I)$ colors of the P0 and P1
Cepheids with the adopted SMC P-C templets leads to the individual
color excesses as a function of period (Fig.~\ref{fig:05}d). The
regression (dashed line), whose equation is given at the bottom of the
panel, is flat and suggests a slightly negative mean color excess
($-0.027\pm0.041\mag$). We interpret this as zero internal absorption.

     The 45 P0 magnitudes in $V$ and $I$ yield, if compared with the
absolute magnitudes from the P-L relation of SMC, individual distances
as a function of $\log P$ (Fig.~\ref{fig:05}e \& f). The dependence of
the distances on $\log P$ is insignificant. Their mean values, read at
the median period, are indicated in the respective panels. The mean
distance from the $V$ and $I$ magnitudes is
$(m-M)^{\rm P0}=24.93\pm0.03$.

     Treating the ten P1 Cepheids in an analogous way gives mean
distances in $V$ and $I$ as shown in panels e \& f of
Fig.~\ref{fig:05} and, if combined, $(m-M)^{\rm P1}=25.01\pm0.02$.

     A number-weighted mean of the P0 distances and the somewhat
larger P1 distances gives $(m-M)^{0}=24.95\pm0.03$ for WLM, which we
adopt.

     The distance modulus of $(m-M)^{0}=25.16$ (normalized to
$(m-M)^{0}_{\rm LMC}=18.52$) of \citeauthor{Pietrzynski:etal:07},
derived from the same data, yet based on Wesenheit pseudo-magnitudes
and using LMC as a templet, is $0.21\mag$ larger than found here.
The modulus of $(m-M)^{0}=24.94\pm0.04$ of \citet{Gieren:etal:08} is
close to the present solution, but the proposed large reddening of
$E(B\!-\!V)=0.08\mag$, implying an {\em internal\/} reddening of
$E(V\!-\!I)=0.03$, would make the WLM Cepheids unusually blue.
The modulus of $24.86\pm0.14$ from four Cepheids with $J$ magnitudes 
\citep{Valcheva:etal:07} is in statistical agreement with the present
value, which is perfectly matched by the value of $24.95\pm0.10$ from
four Cepheids with $3.6$ and $4.5\,\mu$m magnitudes
\citep{Ngeow:etal:09}.

\section{Pegasus = DDO\,216}
\label{sec:5}
The first 6 Cepheids discovered in Pegasus, a peculiar dwarf system,
were measured in $R$ magnitudes by \citet{Hoessel:etal:90} whose
discussion led to a modulus of $(m-M)^{0}=26.22$, which is certainly
too large. \citet{Meschin:etal:09} determined $V$ and $I$ magnitudes
for 18 P0 Cepheids as well as for 8 Cepheids that they identified as
P1 pulsators. These two groups are compared in the following with the
corresponding, calibrated templets of P-C and P-L relations
provided by SMC in Sect.~\ref{sec:2}.  

\begin{figure*}
   \centering
   \resizebox{\figsize\hsize}{!}{\includegraphics{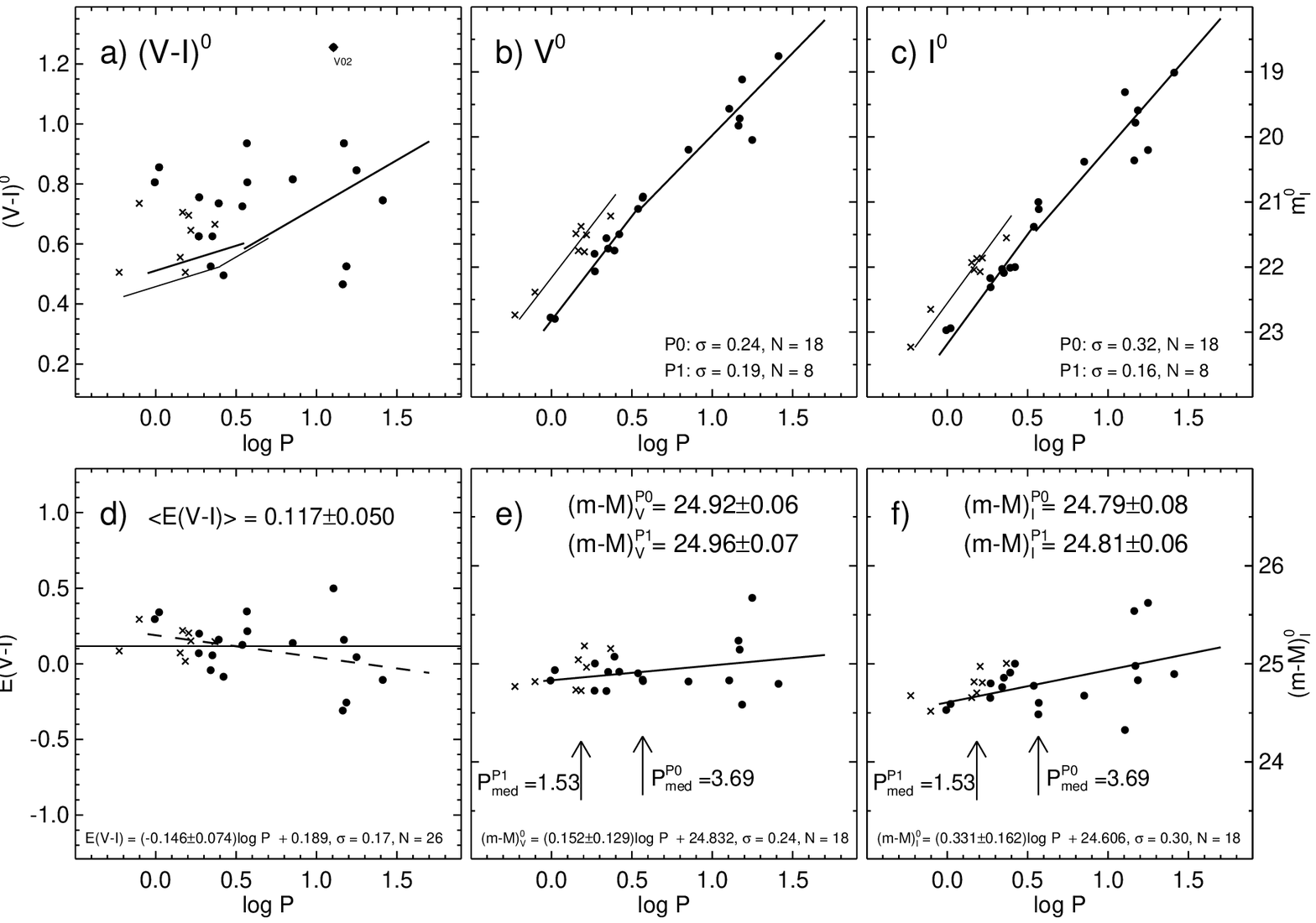}}
   \caption{Same as Fig.~\ref{fig:05}, but for the Cepheids in the
      Pegasus dwarf galaxy.} 
    \label{fig:06}
\end{figure*}

     The P0 Cepheids of \citet{Meschin:etal:09} are shown, after
correction for the Galactic color excess of 0.066
\citep{Schlegel:etal:98} (Schlegel et al. 1998), in the
period-color plane of Fig.~\ref{fig:06}a,
where the P-C relation of SMC is overplotted.
The rather red Cepheids appear to follow a shallow P-C relation, but
the scatter in their data is very large ($0.20\mag$).
The mean color excess derived from Fig.~\ref{fig:06}d of
$E(V\!-\!I)=0.103\pm0.085$, formally decreasing with period, is very
poorly determined. We interpret this as zero absorption for the
moment, but return to this point below.

     The P-L relations in $V$ and $I$ of the P0 and P1 Cepheids are
shown in Fig.~\ref{fig:06}b and c together with the fitted SMC
templets. Eye inspection shows the fits to be excellent. The scatter
in $V$ of the Pegasus Cepheids about the SMC line is essentially the
same as for the SMC Cepheids. The scatter in $I$ is here larger than in
SMC, suggesting that mainly the $I$ magnitudes cause the large
scatter in $(V\!-\!I)$.
Hence, the latter may be affected by observational errors. 

     A comparison of the individual P0 Cepheids with the calibrated
SMC P-L relations leads to the individual distance moduli plotted
versus $\log P$ in Fig.~\ref{fig:06}e \& f. Allowing for their mild
increase with $\log P$ -- the  notorious effect is discussed further
in Sect.~\ref{sec:8} -- we read the mean moduli in $V$ and $I$ at
the median period. The mean distances in $V$ and $I$, as shown in the
respective panels of Fig.~\ref{fig:06}, give a combined distance of
$(m-M)^{P0}=24.86\pm0.06$.
 
     The eight P1 Cepheids, analyzed using the corresponding templet P-L
relations of SMC, lead to the mean moduli in $V$ and $I$ as shown in
Fig.~\ref{fig:06}e \& f and to a combined modulus of 
$(m-M)^{P1}=24.88\pm0.06$ in good agreement with the P0 Cepheids.

     Yet the above assumption of zero internal absorption needs comment. 
The galaxy has a highly variable background, six Cepheids lying on heavy
background and seven are far outlying, but the individual
distances show no correlation with position. This argues against
internal absorption. 
In addition,  the large scatter in the P-L relation for $I$ cannot be
explained by absorption because in this case the scatter in $V$ would
be even larger. The assumption of negligible internal absorption
therefore seems to be justified.

     The overall mean distance, including P0 and P1 Cepheids and $V$
and $I$ colors, is $(m-M)^{0}=24.87\pm0.06$, which we adopt.  

     \citet{Meschin:etal:09} compared only the $V$ magnitudes of
11 longer-period P0 Cepheids with the LMC P-L relations in $V$ of
\citeauthor{STR:04} and \citet{Fouque:etal:07}. This has led to a
somewhat large modulus of $(m-M)^{0}=25.13\pm0.11$.

\section{Sextans A and B}
\label{sec:6}
The two Im dwarf galaxies Sextans~A and Sextans~B lie at the same TRGB
distance \citep[][hereafter TSR\,08b]{TSR:08b} and are separated in
projection by only $\sim\!200\;$kpc. Their recession velocities agree
to within $25\kms$, and they have about the same luminosity and equal,
extremely low metallicities \citep{Sakai:etal:04}. They form a pair,
hence their Cepheid populations are merged here. 

     The first Cepheids in Sextans~A \& B were discussed by
\citet{SC:82,SC:85a}. \citet{Piotto:etal:94} found some additional ones
and gave mean $BVI$ magnitudes for a total of 17 Cepheids (10 in Sextans~A
and 7 in Sextans~B; 4 Cepheids have only $B$ magnitudes). 
The variables P10, P15, and P25 in Sextans~A and P17 in Sextans~B are
probably O1 pulsators.
\citet{Dolphin:etal:03} complemented the sample in Sextans~A with 82
short-period Cepheids with $V$ and $I$ magnitudes, of which 39 are
identified by the authors as P0 pulsators.
In an earlier discussion, we treated the P1 pulsators as P0
Cepheids and concluded that the P-L relations of Sextans~A \& B are much
flatter than those of SMC \citep{ST:08}; an interpretation that cannot
be maintained in the light of the new data.  

\begin{figure*}
   \centering
   \resizebox{\figsize\hsize}{!}{\includegraphics{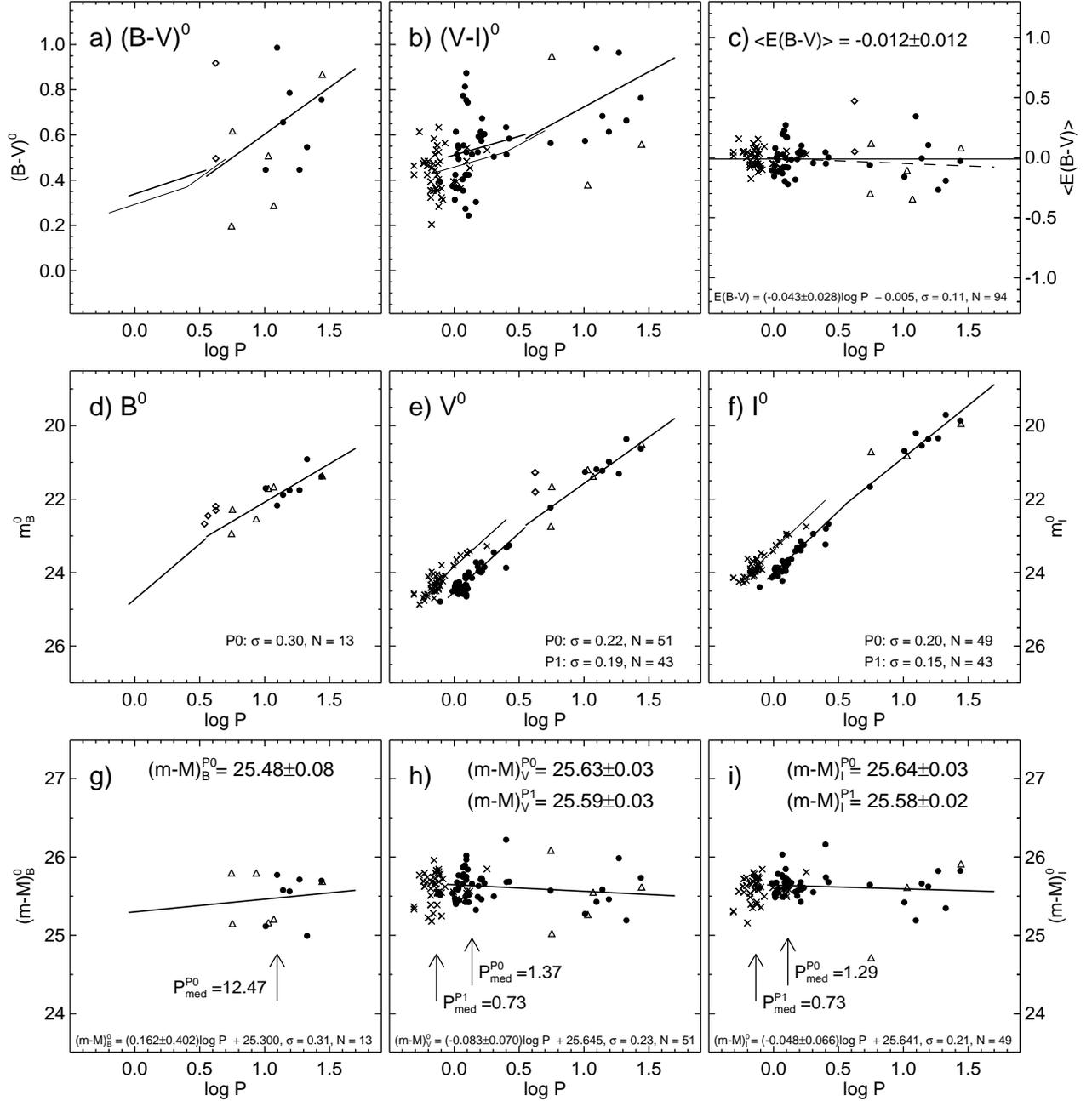}}
   \caption{Analogous to Fig.~\ref{fig:04} but for Cepheids in
     Sextans~A \& B and augmented here with $B$ and $(B\!-\!V)$ data.
} 
    \label{fig:07}
\end{figure*}

     The colors $(B\!-\!V)^{0}$ and $(V\!-\!I)^{0}$ of the
P0 Cepheids are shown as dots (Sextans~A) and open triangles
(Sextans~B) in Fig.~\ref{fig:07}a \& b. Crosses represent P1 Cepheids.
The P-C relations are very poorly determined because of the large scatter.
Yet a comparison of the individual colors with the P-C relation of
SMC leads to $E(B\!-\!V)$ and $E(V\!-\!I)$ excesses. The latter are
converted to $E(B\!-\!V)$ and then averaged. The result is shown in
Fig.~\ref{fig:07}c. The mean excesses depend little on period and
give an overall mean consistent with $E(B\!-\!V)=0.00$.  

     The $B^{0}$, $V^{0}$, and $I^{0}$ magnitudes of the P0 Cepheids 
in Sextans~A (dots) and Sextans~B (open triangles) as well as the P1
Cepheids of Sextans~A (crosses) are plotted versus $\log P$ in
Fig.~\ref{fig:07}d to f. Of the seven Cepheids of Sextans~B, six are taken
as P0 pulsators, the seventh Cepheid may be a P1 pulsator and is omitted.
The templet P-L relations of SMC (from Sect.~\ref{sec:2}) are shown
as heavy lines. They are shifted in magnitude to provide an optimal fit to
the data. The fit is as good as can be expected. The scatter about the
templet lines is about the same as that of the SMC Cepheids.

     The $B^{0}$, $V^{0}$, and $I^{0}$ magnitudes of the P0 Cepheids
are compared with the P-L relations of SMC (Sect.~\ref{sec:2}). This
leads to their individual distances as plotted versus logP in
Fig.~\ref{fig:07}g to i. The distances depend slightly on period, but
the effect is barely significant; we assume, as before, the mean
distance moduli read at the median period of the Cepheids as a
reasonable compromise. The mean distances in $V$ and $I$ agree well.
The mean modulus in $B$ is significantly lower, but is based on only
13 Cepheids. The number-weighted mean over the three colors is
$(m-M)^{\rm P0}=25.60\pm0.05$.

     The 43 P1 Cepheids with $V^{0}$ and $I^{0}$ magnitudes from
\citet{Dolphin:etal:03} are compared with the appropriate P-L relations 
of SMC, shown as thin lines. 
The resulting individual distances are plotted versus $\log P$ in
Fig.~\ref{fig:07}h and i (crosses). They yield a mean modulus of
$(m-M)^{\rm P1}=25.59\pm0.03$ in agreement with the P0 data.
We adopt for the common distance modulus of Sextans~A \& B 
$(m-M)^{0}_{\rm Sex}=25.60\pm0.03$, which is the mean of the 
distances of the P0 and P1 pulsators. -- If Sextans~A and Sextans~B are
treated separately, one finds an overall modulus of
$(m-M)^{0}=25.63\pm0.03$ for Sextans~A  and $(m-M)^{0}=25.53\pm0.10$ for
Sextans~B. The statistical agreement of these two numbers justifies the
combination of the two galaxies into one data set.

     Previous results for the true Cepheid modulus of Sextans~A are
$25.71\pm0.20$ \citep[][including Sextans~B]{Piotto:etal:94} and
$\sim\!25.87\pm0.15$ \citep{Sakai:etal:96}. 
The value of $25.66\pm0.03$ of \citet{Dolphin:etal:03} is based on the
zero-point of SMC, adjusted here to $(m-M)^{0}_{\rm SMC}=18.93$. The
authors converted their $V$ and $I$ magnitudes into Wesenheit
magnitudes, which in this case is not objectionable provided that the
low-metallicity Cepheids in Sextans~A and SMC indeed have identical
colors.

     If the present interpretation is taken at face value, that the
P-C and P-L relations of Sextans~A\&B are at least similar to the ones of
SMC, it follows that below a certain limit of
[O/H]$_{T_{\rm e}}\sim8.0$ the form and the zero-point of these
relations become quite insensitive to metallicity changes of a factor
of $\sim\!3$.

\section{Leo A}
\label{sec:7}
\citet{Dolphin:etal:02} determined the mean $V$ and $R$ magnitudes
for the first 66 unambiguous classical Cepheids in Leo A, of which
the authors classified 19 as P0 and 38 as P1 pulsators.
The P0 Cepheids have periods of between 0.86 and 2.13 days, and the P1
Cepheids of between 0.46 and 1.22 days. These are the shortest-period
Cepheids known.
\citet{Dolphin:etal:02} explain their high frequency -- as
\citet{SC:85b} did before in the case of WLM -- by the metal-dependent
size of the evolutionary loops that feed the instability strip
(see \citealt{Hofmeister:67} and \citealt{Becker:etal:77}).
The absence of Cepheids with longer periods is the result of the
specific star formation rate and the small sample size according to
\citet{Dolphin:etal:02}.  

\begin{figure*}[t]
   \centering
   \resizebox{\figsize\hsize}{!}{\includegraphics{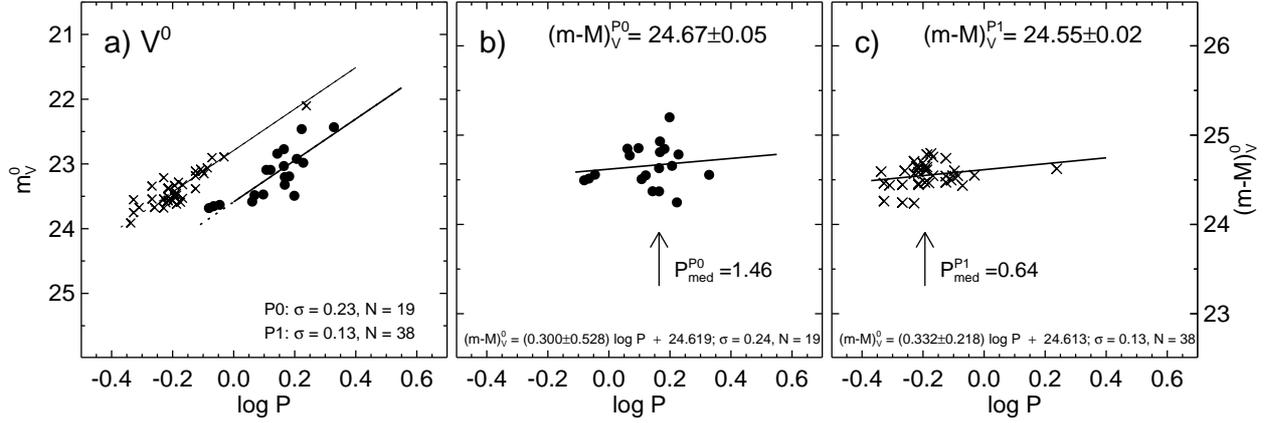}}
   \caption{
    a) The P-L relation in $V$ of the P0 (dots) and P1 (crosses)
    Cepheids in Leo~A. The corresponding, magnitude-shifted relations
    of SMC are shown as heavy lines for P0's and thin lines for P1's.
    b) \& c) The individual distances of the P0 and P1 Cepheids
    plotted versus $\log P$.  
   }
    \label{fig:08}
\end{figure*}

     The P-L relations in $V$, corrected for Galactic absorption, of
the P0 (dots) and P1 (crosses) Cepheids are shown in
Fig.~\ref{fig:08}a. The data can be described well by shifting the templet
relations of SMC in apparent magnitude. We note that the P0 templet is
defined down to only $\sim\!1$ day and the P1 templet down to 0.6 days.
The P-L relations in R are not helpful because  the corresponding data
for SMC are unavailable. It must therefore be assumed that the
internal absorption is negligible, which seems plausible in the light
of the preceding dwarf galaxies.

     The individual distances of the P0 and P1 Cepheids follow
directly from a comparison of their $V^{0}$ magnitudes with the
corresponding calibrated P-L relations of SMC. The dependence of the
resulting distances on $\log P$ is insignificant as seen in 
Fig.~\ref{fig:08}b \& c.
The mean distance modulus, read at median period, is 
$(m-M)^{\rm P0}=24.67\pm0.05$ and
$(m-M)^{\rm P1}=24.55\pm0.02$.
We adopt the number-weighted mean of
$(m-M)^{0}=24.59\pm0.03$.

     \citet[][Table~3]{Dolphin:etal:03} found for P0 and P1 Cepheids
$(m-M)^{0}=24.66$ and 24.54, respectively (adjusted to 
$(m-M)_{\rm SMC}=18.93$) in excellent agreement with the present
result.

\section{Discussion of distances}
\label{sec:8}
The basic assumption of using Cepheids as distance indicators is that
they have the same color (for the determination of the reddening) and
luminosity at a given period. The minimum conditions for this are that
they pulsate in the same mode and that their metallicity is equal.
Whether additional conditions (such as equal He content) should be
fulfilled remains an open question. 

     For every metallicity, one ideally requires a corresponding templet
galaxy with well defined P-C and P-L relations and an independently
known distance. At present, only the Galaxy, LMC, and SMC fulfill these
conditions to serve as templets. Their remaining distance errors of
$0.05\!-\!0.10$ add to the systematic distance error of other galaxies.  

     The original goal of the present paper was to compare the P0
Cepheids of IC\,1613 with those of the SMC because the two galaxies have,
within the errors, the same metallicity. The same holds for the
Cepheids of the Pegasus dwarf galaxy, which were therefore included. On
the basis of the equal metallicities, it is expected that the three
sets of Cepheids define very similar P-C and P-L relations. As shown
above, the expectation is fully confirmed. The next step was to also
include the Cepheids in WLM, the galaxy pair Sextans~A \& B, and Leo~A,
although they are more metal-deficient than SMC by factors of 1.7, 3,
and 4, respectively. In spite of this, no significant differences were
found between their P-C and P-L relations and those of SMC. This
suggests that for very low metallicities ([O/H]$_{T_{\rm e}}\lesssim8.0$) 
even substantial variations in the metallicity have only mild, if any,
effects on the P-C and P-L relations. This is surprising in as much as
the metallicity {\em increase\/} of a factor of 2.3 from SMC to LMC
causes striking differences between the P-C and P-L relations, e.g.\
the overluminosity of the LMC Cepheids, their relative paucity below
$\log P=0.4$, and the break of the LMC relations at $\log P=0.9$
instead of 0.55 in SMC (see Fig.~\ref{fig:10} below).

\begin{table*}
\begin{center}
\caption{Comparison of Cepheid distances with RR~Lyr and TRGB distances.}
\label{tab:04}
\small
\begin{tabular}{lllllll}
\hline
\hline
\noalign{\smallskip}
   &
   \multicolumn{1}{c}{SMC}            &
   \multicolumn{1}{c}{IC 1613}        &
   \multicolumn{1}{c}{WLM}            &
   \multicolumn{1}{c}{Pegasus}        &
   \multicolumn{1}{c}{Sextans~A \& B} &
   \multicolumn{1}{c}{Leo~A}          \\
\noalign{\smallskip}
\hline
\noalign{\smallskip}
Cep P0             & $18.93^{1)}$            & $24.32\pm0.02$            & $24.93\pm0.03$            & $24.86\pm0.06$            & $25.60\pm0.05$            & $24.67\pm0.05$            \\
Cep P1             & $18.93^{1)}$            & $24.38\pm0.03$            & $25.01\pm0.02$            & $24.88\pm0.06$            & $25.59\pm0.03$            & $24.55\pm0.02$            \\
{\bf Cep adopted}  & \boldmath{$18.93^{1)}$} & \boldmath{$24.34\pm0.03$} & \boldmath{$24.95\pm0.03$} & \boldmath{$24.87\pm0.06$} & \boldmath{$25.60\pm0.04$} & \boldmath{$24.59\pm0.03$} \\
RR~Lyr             & $18.98$                 & $24.35$                   & \nodata                   & \nodata                   & \nodata                   & $24.54$                   \\
TRGB               & $19.00$                 & $24.32$                   & $24.90$                   & $24.84$                   & $25.72$                   & $24.57$                   \\
\noalign{\smallskip}
$\Delta(m-M)^{0}$ & \nodata                  & $0.02$                    & $0.05$                   & $0.03$                    & $-0.12$                   & $0.02$                     \\ 
\noalign{\smallskip}
\hline
\end{tabular}
\end{center}
\tablefoot{%
$^{1)}$used as calibrator (see \citeauthor{STR:09})     
}
\end{table*}

     The near equality of the properties of the Cepheids in
the present sample suggests that the errors of the adopted distances
in Table~\ref{tab:04} are caused mainly by statistics and the
zero-point error of the SMC. 

     Yet the slopes of the P-C and P-L relations of two galaxies are
hardly ever identical -- be it for intrinsic or statistical
reasons. Any slope difference of the P-C relations of  the galaxy under
investigation and the galaxy used as a templet will lead to color
excesses that vary with period. 
This is not a serious problem in the present case because all excesses
are vanishingly small.
   
     In addition, any slope differences between the P-L relation of the
galaxy under investigation and the templet P-L relation will cause 
the distances of individual Cepheids to depend on period. 
If the former has the slope $p_{1}\pm\epsilon_{1}$ and the latter an
observed slope of $p_{2}\pm\epsilon_{2}$, then the slope of the $(m-M)
- \log P$ relation will have the slope $\pi=p_{1}-p_{2}$ with 
a seemingly small error of $\pm\epsilon_{2}$ because the templet is
assumed to be error-free. In principle, the problem cannot be solved
because the slope differences may be real. Nevertheless, if we assume
that the majority of the sample Cepheids follow the templet P-L
relation, the best mean distance is read at their median period. This
is justified, however, only if the templet P-L relation is well defined
at this period, which is the case for the present sample of galaxies.
The problem is aggravated whenever the centers of weight of the test
and templet Cepheids lie at different periods, as happens frequently
for more distant galaxies that are biased toward longer-period
Cepheids. All published Cepheid distances we tested exhibit the
dependence of the modulus on $\log P$ \citep[see e.g.][]{Saha:etal:06}, 
which adds to the external error in the Cepheid distances more than is
generally acknowledged.

     In view of the possible remaining systematic errors, it is
important to test the derived Cepheid distances in the light of
{\em independent\/} distance determinations. The most reliable
alternative distance indicators are RR~Lyr stars and the tip of the
red-giant branch (TRGB). The available data are compiled in
Table~\ref{tab:04}.  
The P0, P1, and adopted Cepheid distances and their estimated internal
errors are repeated from Sects.~\ref{sec:3}-\ref{sec:7}. The RR~Lyr
distances are taken from \citeauthor*{TSR:08a}. The TRGB distances
were compiled by \citeauthor*{TSR:08b} and augmented here by some
additional sources   
(\citealt{Lee:etal:93};
 \citealt{Aparicio:94};
 \citealt{Minniti:Zijlstra:97};
 \citealt{Mendez:etal:02};
 \citealt{Dolphin:etal:03}; 
 \citealt{Tully:etal:06};
 \citealt{Meschin:etal:09};
 \citealt{Dalcanton:etal:09}).
In total, 22 measurements of the TRGB were used for the five
galaxies in Table~\ref{tab:04}. They are homogenized to a common
zero-point of $M^{0}_{\rm TRGB}=-4.05$ as determined from 19 RR~Lyr
distances (\citeauthor*{TSR:08a}) and independently confirmed by
\citet{Rizzi:etal:07}. Although theory predicts that the TRGB depends
somewhat on metallicity, the sign of the correction remains under
discussion. 

     The comparison in Table~\ref{tab:04} is surprisingly good. P0 and
P1 distances agree on average to within $0.01\pm0.04\mag$. The two
RR~Lyr distances deviate from the corresponding Cepheid distances by
only $\pm0.01$. The mean difference between the Cepheid and TRGB
distances is $\langle\Delta(m-M)\rangle=0.00\pm0.03$. The null result
not only supports the adopted zero-point distance of SMC, but provides
a consistency check of the Population~I and Population~II distance
scales. The rms of the distance differences is
$\sigma_{(m-M)}=0.07\mag$. Even if one allows a value as low as 0.05 
for the random error in the TRGB distances , the random external error
in the Cepheid distances is not more than $0.05\mag$. 

\begin{figure*}
   \centering
   \resizebox{\figsize\hsize}{!}{\includegraphics{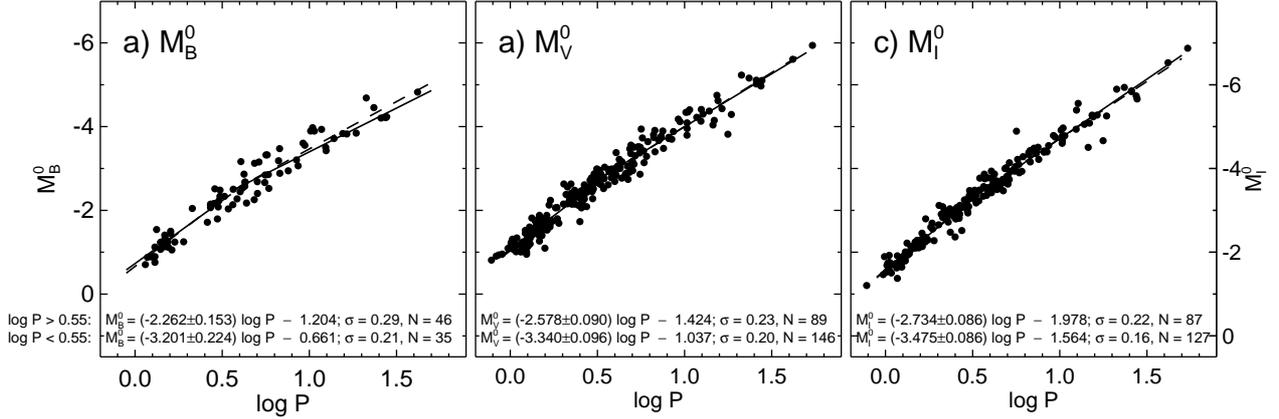}}
   \caption{Composite P-L relations in $M_{B}$, $M_{V}$, and
   $M_{I}$ of the P0 Cepheids in IC\,1613, WLM, Sextans A\&B, the Pegasus
   dwarf, and Leo~A  using their adopted distances.
   The dashed lines are regressions to the Cepheids with $\log
   P\gtrless 0.55$; their equations are given at the bottom of each panel.
   The full lines are the corresponding P-L relations of SMC. The two
   lines in each panel are almost congruent.  
}   
    \label{fig:09}
\end{figure*}

\section{Results and conclusions}
\label{sec:9}
The P0 Cepheids in IC\,1613, WLM, Pegasus, Sextans~A \& B, and Leo~A
(excluding SMC) are combined into composite P-L relations in $B$, $V$,
and $I$, adopting the respective Cepheid distances derived in
Sects.~\ref{sec:3}-\ref{sec:7} (Fig.~\ref{fig:09}). The resulting P-L
relations, whose equations are at the foot of Fig.~\ref{fig:09}, are
indistinguishable from those for SMC. Over the period interval of
$0.2<\log P<1.2$, the P-L relations of SMC and the five sample
galaxies agree to better than $0.02\mag$ in $V$ and $I$. In $B$, with
fewer variables the agreement is not quite as good. In addition, the P1
Cepheids define closely agreeing P-L relations for SMC and the
combined sample of five galaxies. 
This proves -- in agreement with our prediction -- that the P-L
relations of SMC hold for the equally metal-poor galaxies IC\,1613,
WLM, and Pegasus and even for the still more metal-poor Sextans~A \& B
and probably also for Leo~A. (In the case of Leo~A, the comparison is
restricted to Cepheids with $\log P<0.4$). The low-metallicity
galaxies are therefore part of a family with (nearly) equal P-L
relations. This holds of course also for the P-C relations, which are
nothing else but the difference of the respective P-L relations. 

     Cepheids of higher metallicity, such as those in LMC and the Galactic
Cepheids in the solar neighborhood, have distinctly different P-L and
P-C relations. For convenience, the coefficients of the relevant
equations for the P0 Cepheids are compiled here in Table~\ref{tab:05}
following the scheme $x=a\log P + b$. The equations for SMC and LMC
follow from Sect.~\ref{sec:2}. The Galactic equations come from
\citeauthor{TSR:03} and the revision in \citeauthor{STR:04}.

\begin{table*}
\begin{center}
\caption{Coefficients of the relevant P-C and P-L relations for
  P0 Cepheids. Slope coefficients that agree to within $1\sigma$ are
  underlined.}
\label{tab:05}
\footnotesize
\begin{tabular}{lrrrrcrrrrcrr}
\hline
\hline
\noalign{\smallskip}
   &
   \multicolumn{4}{c}{SMC$^{1)}$} & &
   \multicolumn{4}{c}{LMC$^{2)}$} & &
   \multicolumn{2}{c}{Galaxy} \\
   &
   \multicolumn{4}{c}{[O/H]=7.98} & &
   \multicolumn{4}{c}{[O/H]=8.36} & &
   \multicolumn{2}{c}{[O/H]=8.62} \\
   &
   \multicolumn{2}{c}{$\log P<0.55$} &
   \multicolumn{2}{c}{$\log P>0.55$} & &
   \multicolumn{2}{c}{$\log P<0.9$} &
   \multicolumn{2}{c}{$\log P>0.9$}     & &
   \multicolumn{2}{c}{} \\
   &
   \multicolumn{1}{c}{$a$} &
   \multicolumn{1}{c}{$b$} &
   \multicolumn{1}{c}{$a$} &
   \multicolumn{1}{c}{$b$} & &
   \multicolumn{1}{c}{$a$} &
   \multicolumn{1}{c}{$b$} &
   \multicolumn{1}{c}{$a$} &
   \multicolumn{1}{c}{$b$} & &
   \multicolumn{1}{c}{$a$} &
   \multicolumn{1}{c}{$b$} \\
\noalign{\smallskip}
\hline
\noalign{\smallskip}
$(B\!-\!V)^{0}$ &    $0.191$ &    $0.339$ &    \underline{$0.415$} &    $0.188$ &&    $0.306$ &    $0.330$ &    \underline{$0.435$} &    $0.199$ &&    $0.366$ &    $0.361$ \\[-2pt]
                & $\pm0.021$ & $\pm0.005$ & $\pm0.020$ & $\pm0.018$ && $\pm0.020$ & $\pm0.012$ & $\pm0.029$ & $\pm0.036$ && $\pm0.015$ & $\pm0.013$ \\[2pt]
$(V\!-\!I)^{0}$ &    $0.166$ &    $0.511$ &    \underline{$0.311$} &    $0.413$ &&    $0.201$ &    $0.474$ &    \underline{$0.345$} &    $0.331$ &&    $0.256$ &    $0.497$ \\[-2pt]
                & $\pm0.018$ & $\pm0.005$ & $\pm0.016$ & $\pm0.014$ && $\pm0.017$ & $\pm0.010$ & $\pm0.024$ & $\pm0.030$ && $\pm0.017$ & $\pm0.016$ \\[2pt]
    $M_{B}^{0}$ &   $-3.007$ &   $-0.728$ &   \underline{$-2.091$} &   $-1.306$ &&   $-2.491$ &   $-1.083$ &   \underline{$-2.021$} &   $-1.576$ &&   $-2.692$ &   $-0.575$ \\[-2pt]
                & $\pm0.076$ & $\pm0.022$ & $\pm0.071$ & $\pm0.063$ && $\pm0.067$ & $\pm0.040$ & $\pm0.100$ & $\pm0.123$ && $\pm0.093$ & $\pm0.107$ \\[2pt]
    $M_{V}^{0}$ &   $-3.203$ &   $-1.071$ &   \underline{$-2.531$} &   $-1.466$ &&   $-2.787$ &   $-1.414$ &   \underline{$-2.505$} &   $-1.713$ &&   $-3.087$ &   $-0.914$ \\[-2pt]
                & $\pm0.060$ & $\pm0.018$ & $\pm0.056$ & $\pm0.050$ && $\pm0.048$ & $\pm0.029$ & $\pm0.074$ & $\pm0.091$ && $\pm0.085$ & $\pm0.098$ \\[2pt]
    $M_{I}^{0}$ &   \underline{$-3.374$} &   $-1.577$ &   \underline{$-2.842$} &   $-1.872$ &&   $-3.008$ &   $-1.880$ &   \underline{$-2.812$} &   $-2.076$ &&   \underline{$-3.348$} &   $-1.429$ \\[-2pt]
                & $\pm0.046$ & $\pm0.013$ & $\pm0.043$ & $\pm0.038$ && $\pm0.032$ & $\pm0.019$ & $\pm0.057$ & $\pm0.069$ && $\pm0.083$ & $\pm0.097$ \\
\noalign{\smallskip}
\hline
\end{tabular}
\end{center}
\tablefoot{%
   $^{1)}$ adopted at $(m-M)^{0}_{\rm SMC}=18.93$ (\citeauthor*{TSR:08b}, Table~6);  
   $^{2)}$ adopted at $(m-M)^{0}_{\rm LMC}=18.52$ (\citeauthor*{TSR:08b}, Table~7)
}    
\end{table*}

     The steep slopes of the Galactic P-L relations from Paper~I and II
corresponds to data for Cepheids in Galactic clusters and OB
associations \citep{Feast:99} as well as Baade-Wesselink-Becker
distances by \citet{Fouque:etal:03}  and \citet{Barnes:etal:03}, the
two fully independent methods leading to the same result.
Criticism of the result was discussed by \citeauthor*{TSR:08b}.
The P-L relations of metal-rich Cepheids will be discussed in more
detail in a forthcoming paper; it is possible that they experience a
break at long periods ($\log P\ga1.6$), but this is irrelevant here.

     The observed P-L relations of LMC are closely matched -- including
the break at $\sim\!10^{\rm d}$ -- by theoretical P-L relations based
on pulsation models \citep{Marconi:etal:05}. The same models do not
show a break at higher metallicities in agreement with the Galactic
P-L relations adopted here.

     \citet{Marconi:etal:10} also derived theoretical P-L
relations for ultra-low metallicities. They have no break and are
somewhat flatter than in SMC up to $\log P=0.55$, but are much steeper
beyond that point. The comparison may not be justified because the
adopted metallicity ([O/H]$\sim 7.0$) is lower than in SMC and even
Leo~A.

\begin{figure*}
   \centering
   \resizebox{0.7\hsize}{!}{\includegraphics{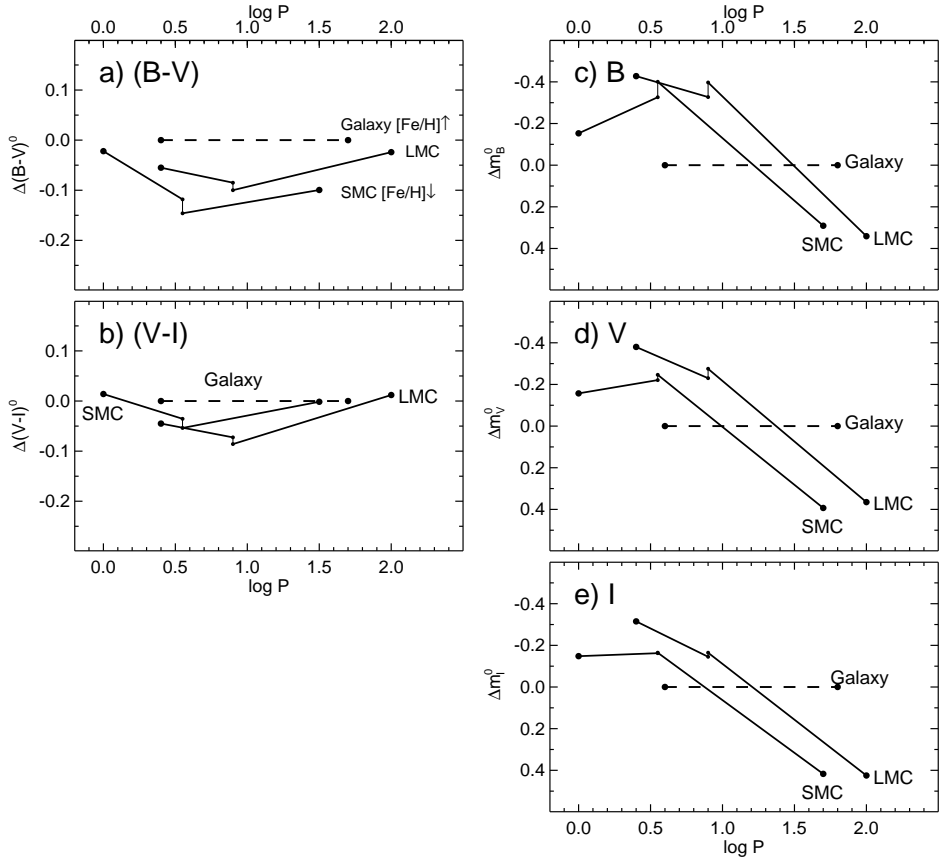}}
   \caption{a) The P-C relation in $(B\!-\!V)$ of P0 Cepheids
   in both the very metal-poor SMC and the relatively metal-poor LMC
   {\em relative\/} to the metal-rich Solar neighborhood. 
   b) Same as a) but for $(V\!-\!I)$. 
   c) - e) The P-L relations in $B$, $V$, and $I$, respectively, of
   the P0 Cepheids in SMC and LMC {\em relative\/} to the Solar
   neighborhood. The artificial spikes of the relations are due to
   statistical errors of the fits below and above the break. The lines
   are only drawn over the period range where they are well defined by
   observations.} 
    \label{fig:10}
\end{figure*}

     To illustrate the difference between the Cepheids in SMC, LMC,
and the solar neighborhood, the P-C and P-L relations of SMC and LMC
are plotted {\em relative\/} to those of the Galaxy in
Fig.~\ref{fig:10}. In each panel, the Galactic relations are taken as
reference and the {\em differences\/} in color and absolute magnitude
of the Cepheids in the other two galaxies are shown as a function of
$\log P$ (in the sense $x_{\rm LMC/SMC}-x_{\rm Galaxy}$). 

     As seen in Fig.~\ref{fig:10}a the LMC Cepheids are bluer in
$(B\!-\!V)$ than their Galactic counterparts by up to $0.09\mag$ at
$\log P=0.9$. The color difference is even larger between SMC and the
Galaxy, i.e.\ $0.13\mag$ at $\log P=0.55$. The red color of the
Galactic Cepheids is due to their lower temperature and the blanketing
effect of the metal lines (see \citeauthor{STR:04}).

     The color differences in $(V\!-\!I)$ between the three galaxies
in Fig.~\ref{fig:10}b are smaller. The LMC Cepheids are bluer than in
the Galaxy by up to $0.08\mag$. Unexpectedly, the SMC Cepheids are
redder than in LMC, yet still bluer than those in the Galaxy by a
marginal amount of $0.04\mag$ or less, depending on period.

     The P-L relations in $B$, $V$, and $I$ of LMC and SMC are plotted
relative to the Galactic P-L relations in Fig.~\ref{fig:10}c-e. The
relations of LMC and SMC have similar characteristics and differ
mainly in the zero-point, but they are both much flatter than in the
Galaxy beyond the break point. At $\log P=0.6$, LMC and SMC Cepheids
are respectively brighter by $0.39\mag$ and $0.37\mag$ than in the
Galaxy, whereas at $\log P = 1.7$ they are fainter by 0.14 and
$0.29\mag$, respectively. LMC Cepheids are brighter than in SMC by
$0.15-0.20\mag$, somewhat depending on period. 
The significant luminosity differences of the Cepheids in the three
galaxies cannot be explained by errors in the adopted distances, which
are on the order of $0.1\mag$. In addition, it is impossible to explain
the different slopes of the P-L relations by distance errors.

     We note that some of the {\em slopes\/} of the P-L
relations in Table~\ref{tab:05} show striking agreement. The slopes of
SMC and LMC are essentially identical in $B$, $V$, and $I$ above the
break points, and the short-period SMC P-L relation in $I$ has the
same slope as Galactic Cepheids. In addition, the slopes of the P-C relations
of SMC and LMC are the same to within $\sim\!1\sigma$ for $\log P>0.9$.

     The Cepheids designated here as low-metallicity objects comprise in
fact a wide metallicity range of $8.0>$[O/H]$T_{\rm e}>7.4$. Their
very similar P-L relations imply that they are quite insensitive at
these low levels to metallicity changes. This is in sharp contrast to
more metal-rich Cepheids where a change of only $\Delta$[O/H]$T_{\rm e}=0.26$  
causes the pronounced differences between the LMC and Galactic P-L
relations.

     The use of Cepheids as distance indicators has been extended here to
include fundamental-mode (P0) and first-overtone (P1) Cepheids with
the shortest periods known. Among the known Cepheid population of the SMC,
47\% of the P0 pulsators have periods less than $\log P=0.4$, extending
down to $\log P=0.0$, and 37\% are P1 pulsators with periods down to
$\log P=-0.2$. The large number of these additional Cepheids makes them
indispensable for accurate distance determinations. The distances
derived here agree with independent RR~Lyr and TRGB distances to
within a few $0.01\mag$.

\begin{acknowledgements}
We thank the referee, Dr. Fran\c{c}oise Combes, for very helpful comments.
\end{acknowledgements}




\begin{thebibliography}{}
%
\bibitem[Antonello et~al.(2006)]{Antonello:etal:06}
   Antonello, E., Fossati, L., Fugazza, D., Mantegazza, L., \& Gieren, W. 2006,
   A\&A, 445, 901 (IC\,1613)
%
\bibitem[Aparicio(1994)]{Aparicio:94}
   Aparicio, J.~M. 1994,
   ApJ, 437, L27
%
\bibitem[Barnes et~al.(2003)]{Barnes:etal:03}
   Barnes, T., Jeffreys, W., Berger, J., et~al. 2003, 
   ApJ, 592, 539
%
\bibitem[Becker et~al.(1977)]{Becker:etal:77}
   Becker, S.~A., Iben, I., Jr., \& Tuggle, R.~S. 1977, 
   ApJ, 218, 633
%
\bibitem[Bernard et~al.(2010)]{Bernard:etal:10}
   Bernard, E.~J., Monelli, M., Gallart, C., et~al. 2010, 
   ApJ, 712, 1259 (IC\,1613)
%
\bibitem[Dalcanton et~al.(2009)]{Dalcanton:etal:09}
   Dalcanton, J.~J., Williams, B.~F., Seth, A.~C.,  et~al. 2009, 
   ApJS, 183, 67 
%
\bibitem[de~Vaucouleurs(1978)]{deVaucouleurs:78}
   de~Vaucouleurs, G. 1978,
   ApJ 223, 730
%
\bibitem[Dolphin et~al.(2002)]{Dolphin:etal:02}
   Dolphin, A.~E., Saha, A., Claver, C. et~al. 2002, 
   AJ, 123, 3154 (Leo~A)
%
\bibitem[Dolphin et~al.(2001)]{Dolphin:etal:01}
   Dolphin, A.~E., Saha, A., Skillman, E.~D., et~al. 2001, 
   ApJ, 550, 554 (IC\,1613)
%
\bibitem[Dolphin et~al.(2003)]{Dolphin:etal:03}
   Dolphin, A.~E., Saha, A., Skillman, E.~D., et~al. 2003, 
   AJ, 125, 1261 (Sextans~A)
%
\bibitem[Feast(1999)]{Feast:99}
   Feast, M.~W. 1999,
   PASP, 111, 775 
%
\bibitem[Fiorentino et~al.(2010)]{Fiorentino:etal:10}
   Fiorentino, G., Contreras Ramos, R., Clementini, G., et~al. 2010, 
   ApJ, 711, 808 
%
\bibitem[Fouqu{\'e} et~al.(2003)]{Fouque:etal:03}
   Fouqu{\'e}, P., Storm, J., \& Gieren, W. 2003,
   Lect. Notes Phys., 635, 21
%
\bibitem[Fouqu{\'e} et~al.(2007)]{Fouque:etal:07}
   Fouqu{\'e}, P., Arriagada, P., Storm, J., et~al. 2007, 
   A\&A, 476, 73 
%
\bibitem[Freedman(1988a)]{Freedman:88}
   Freedman, W.~L. 1988a,
   AJ, 96, 1248 (IC\,1613)
%
\bibitem[Freedman(1988b)]{Freedman:88b}
   Freedman, W.~L. 1988b,
   ApJ, 326, 691
%
\bibitem[Freedman et~al.(2009)]{Freedman:etal:09}
   Freedman, W.~L., Rigby, J., Madore, B.~F., et~al. 2009, 
   ApJ, 695, 996
%
\bibitem[Gieren et~al.(2008)]{Gieren:etal:08}
   Gieren, W., Pietrzy{\'n}ski, G., Szewczyk, O., et~al. 2008, 
   ApJ, 683, 611
%
\bibitem[Hoessel et~al.(1990)]{Hoessel:etal:90}
   Hoessel, J.~G., Abbott, M.~J., Saha, A., Mossman, A.~E., \&
   Danielson, G. E. 1990,  
   AJ, 100, 1151
%
\bibitem[Hofmeister(1967)]{Hofmeister:67}
   Hofmeister, E. 1967,
   Z. Astrophys., 65, 194
%
\bibitem[Kanbur \& Ngeow(2004)]{Kanbur:Ngeow:04}
   Kanbur, S.~M., \& Ngeow, C.-C. 2004,
   MNRAS, 350, 962
%
\bibitem[Kanbur et~al.(2007)]{Kanbur:etal:07}
   Kanbur, S.~M., Ngeow, C.-C., Nanthakumar, A., \& Stevens, R. 2007, 
   PASP, 119, 512
%
\bibitem[Koen \& Siluyele(2007)]{Koen:Siluyele:07}
   Koen, C., \& Siluyele, I. 2007,
   MNRAS, 377, 1281
%
\bibitem[Lee et~al.(1993)]{Lee:etal:93}
   Lee, M.~G., Freedman, W.~L., \& Madore, B.~F. 1993, 
   ApJ, 417, 553 
%
\bibitem[Macri et~al.(2001)]{Macri:etal:01}
   Macri, L.~M., Calzetti, D., Freedman, W.~L., et~al. 2001, 
   ApJ, 549, 721
%
\bibitem[Madore \& Freedman(1991)]{Madore:Freedman:91}
   Madore, B.~F., \& Freedman, W.~L. 1991,
   PASP, 103, 933
%
\bibitem[Marconi et~al.(2005)]{Marconi:etal:05}
   Marconi, M., Musella, I., \& Fiorentino, G. 2005,
   ApJ, 632, 590
%
\bibitem[Marconi et~al.(2010)]{Marconi:etal:10}
   Marconi, M., Musella, I., Fiorentino, G., et~al. 2010, 
   ApJ, 713, 615
%
\bibitem[McAlary et al.(1984)]{McAlary:etal:84}
   McAlary, C.~W., Madore, B.~F., \& Davis, L.~E. 1984,
   ApJ, 276, 487
%
\bibitem[M{\'e}ndez et~al.(2002)]{Mendez:etal:02}
   M{\'e}ndez, B., Davis, M., Moustakas, J., et~al. 2002, 
   AJ, 124, 213
%
\bibitem[Meschin et~al.(2009)]{Meschin:etal:09}
   Meschin, I., Gallart, C., Aparicio, A., Cassisi, S., \& Rosenberg, A. 2009,
   AJ, 137, 3619 (Pegasus)
%
\bibitem[Minniti, \& Zijlstra(1997)]{Minniti:Zijlstra:97}
   Minniti, D., \& Zijlstra, A.~A., 1997,
   AJ, 114, 147
%
\bibitem[Ngeow et~al.(2009)]{Ngeow:etal:09}
   Ngeow, C.-C., Kanbur, S.~M., Ghobrial, L., Neilson, H., \& Macri, L. 2009, 
   in Stellar Pulsation: Challenges for Theory and Observation,
   ed. J.~A. Guzik \& P.~A. Brodley (College Park: AIP), 37
%
\bibitem[Ngeow et~al.(2005)]{Ngeow:etal:05}
   Ngeow, C.-C., Kanbur, S.~M., Nikolaev, S., et~al. 2005, 
   MNRAS, 363, 831
%
\bibitem[Pietrzy{\'n}ski et~al.(2006)]{Pietrzynski:etal:06}
   Pietrzy{\'n}ski, G., Gieren, W., Udalski, A., et~al. 2006, 
   ApJ, 642, 216
%
\bibitem[Pietrzy{\'n}ski et~al.(2007)Pietrzy{\'n}ski's et~al.]{Pietrzynski:etal:07}
   Pietrzy{\'n}ski, G., Gieren, W., Udalski, A., et~al. 2007, 
   AJ, 134, 594 (WLM)
%
\bibitem[Piotto et~al.(1994)]{Piotto:etal:94}
   Piotto, G., Capaccioli, M., \& Pellegrini, C. 1994,
   A\&A, 287, 371 (Sextans A\&B)
%
\bibitem[Rizzi et~al.(2007)]{Rizzi:etal:07}
   Rizzi, L., Tully, R.~B., Makarov, D., et~al. 2007, 
   ApJ, 661, 815
%
\bibitem[Saha et~al.(2006)]{Saha:etal:06}
   Saha, A., Thim, F., Tammann, G.~A., Reindl, B., \& Sandage, A. 2006,
   ApJS, 165, 108
%
\bibitem[Sakai et~al.(2004)]{Sakai:etal:04}
   Sakai, S., Ferrarese, L., Kennicutt, R.~C., \& Saha, A. 2004,
   ApJ, 608, 42
%
\bibitem[Sakai et~al.(1996)]{Sakai:etal:96}
   Sakai, S., Madore, B.~F., \& Freedman, W.~L. 1996,
   ApJ, 461, 713
%
\bibitem[Sandage(1971)]{Sandage:71}
   Sandage, A. 1971,
   ApJ, 166, 13
%
\bibitem[Sandage \& Carlson(1982)]{SC:82}
   Sandage, A., \& Carlson, G. 1982,
   ApJ, 258, 439
%
\bibitem[Sandage \& Carlson(1985a)]{SC:85a}
   Sandage, A., \& Carlson, G. 1985a,
   AJ, 90, 1019
%
\bibitem[Sandage \& Carlson(1985b)]{SC:85b}
   Sandage, A., \& Carlson, G. 1985b,
   AJ, 90, 1464
%
\bibitem[Sandage \& Tammann(2008)ST\,08]{ST:08}
   Sandage, A., \& Tammann, G.~A. 2008,
   ApJ, 686, 779
%
\bibitem[Paper~II(2004)Sandage et~al.]{STR:04}
   Sandage, A., Tammann, G.~A., \& Reindl, B. 2004,
   A\&A, 424, 43  (Paper~II)
%
\bibitem[Paper~III(2009)Sandage et~al.]{STR:09}
   Sandage, A., Tammann, G.~A., \& Reindl, B. 2009,
   A\&A, 493, 471  (Paper~III)
%
\bibitem[Schlegel et~al.(1998)]{Schlegel:etal:98}
   Schlegel, D., Finkbeiner, D., \& Davis, M. 1998,
   ApJ, 500, 525
%
\bibitem[Skillman \& Kennicutt(1993)]{Skillman:Kennicutt:93}
   Skillman, E.~D., \& Kennicutt, R.~C. 1993,
   ApJ, 411, 655
%
\bibitem[Skillman et~al.(1989)]{Skillman:etal:89}
   Skillman, E.~D., Kennicutt, R.~C., \& Hodge, P.~W.  1989,
   ApJ, 347, 875
%
\bibitem[Skillman et~al.(1997)]{Skillman:etal:97}
   Skillman, E.~D.,  Bomans, D.~J., \& Kobulnicky, H.~A. 1997,
   ApJ, 474, 205
%
\bibitem[Soszy{\'n}ski et~al.(2010)]{Soszynski:etal:10}
   Soszy{\'n}ski, I., Poleski, R., Udalski, A., et~al. 2010, 
   AcA, 60, 17 
%
\bibitem[Tammann \& Reindl(2002)]{Tammann:Reindl:02}
   Tammann, G.~A., \&  Reindl, B. 2002,
   Ap\&SS, 280, 165

\bibitem[Tammann et~al.(2002)]{Tammann:etal:02}
   Tammann, G.~A., Reindl, B., Thim, F., Saha, A., \& Sandage, A. 2002,
   in A New Era in Cosmology,
   ed. T.~Shanks, \& N.~Metcalfe,
   ASP Conf. Ser., 283, 258
%
\bibitem[Paper~I(2003)Tammann et~al.]{TSR:03}
   Tammann, G.~A., Sandage, A., \& Reindl, B. 2003, 
   A\&A, 404, 423 (Paper~I)
%
\bibitem[Tammann et~al.(2008a)TSR\,08a]{TSR:08a}
   Tammann, G.~A., Sandage, A., \& Reindl, B. 2008a,
   ApJ, 679, 52  (TSR\,08a) 
%
\bibitem[Tammann et~al.(2008b)TSR\,08b]{TSR:08b}
   Tammann, G.~A., Sandage, A., \& Reindl, B. 2008b,
   A\&ARv, 15, 289  (TSR\,08b)
%
\bibitem[Tully et~al.(2006)]{Tully:etal:06}
   Tully, R.~B., Rizzi, L., Dolphin, A.~E., et~al. 2006, 
   AJ, 132, 729
%
\bibitem[Udalski et~al.(1999a)]{Udalski:etal:99a}
   Udalski, A., Soszynski, I., Szymanski, M., et~al. 1999a, 
   AcA, 49, 223 (LMC)
%
\bibitem[Udalski et~al.(1999b)]{Udalski:etal:99b}
   Udalski, A., Soszynski, I., Szymanski, M., et~al. 1999b, 
   AcA, 49, 437 (SMC)
%
\bibitem[Udalski et~al.(2001)]{Udalski:etal:01}
   Udalski, A., Wyrzykowski, L., Pietrzynski, G., et~al., 2001, 
   AcA 51, 221 (IC\,1613)
%
\bibitem[Valcheva et~al.(2007)]{Valcheva:etal:07} 
   Valcheva, A.~T., Ivanov, V.~D., Ovcharov, E.~P., \& Nedialkov, P.~L. 2007, 
   A\&A, 466, 501
%
\bibitem[van~den~Bergh(1968)]{vandenBergh:68} 
   van~den~Bergh, S. 1968, 
   R.~A.~S.~C. Jour., 62, 145
%
\bibitem[van~Zee et~al.(2006)]{vanZee:etal:06} 
   van Zee, L., Skillman, E.~D., \& Haynes, M.~P. 2006, 
   ApJ, 637, 269
%
\bibitem[Zaritsky et~al.(1994)]{Zaritsky:etal:94}
   Zaritsky, D., Kennicutt, R.~C., \& Huchra, J.~P. 1994, 
   ApJ, 420, 87
%
\end{thebibliography}
\end{document}